\documentclass[12pt]{article}
 
\pdfoutput=1

 \usepackage{amsmath, amssymb, amscd, amsthm, amsfonts}
 \usepackage{physics}
 \usepackage{graphicx}
 \usepackage{braket}
  \usepackage{bbm}
 \usepackage{subfigure}
 \usepackage{tikz}
 \usetikzlibrary{decorations.pathmorphing}
 \usepackage{comment}
 \usepackage{soul}
 \usepackage{multirow}
 \usepackage{booktabs}
 \usepackage{tikz}
 \usetikzlibrary{arrows.meta, decorations.pathreplacing}

\usepackage{verbatim}

\usepackage{amsmath,amssymb,amsfonts}
\usepackage{psfrag}
\usepackage{footmisc}
\usepackage{url}
\usepackage{mathtools}
\usepackage{color}
\usepackage[dvipsnames]{xcolor}

\usepackage{tikz}
    \usepackage{amssymb,amsfonts,amsmath}
    \usepackage{tkz-euclide}
        \usetikzlibrary{arrows,calc,patterns}
\usepackage{pgfplots}

\definecolor{darkblue}{rgb}{0.1,0.1,.7}
\definecolor{darkgreen}{rgb}{0.1,0.6,.1}
\definecolor{purple}{rgb}{0.6,0,0.6}
\definecolor{orange}{rgb}{0.9,0.6,0}
\definecolor{lightred}{rgb}{0.85,0.35,0.35}
\usepackage[colorlinks, linkcolor=darkblue, citecolor=darkblue, urlcolor=darkblue, linktocpage]{hyperref} 
\usepackage[square, comma, compress,numbers]{natbib}
\usepackage[]{amsmath}
\usepackage[]{graphicx}
\usepackage[]{latexsym}
\usepackage[utf8]{inputenc}
\usepackage{geometry}
\usepackage{amscd}
\usepackage[all,cmtip]{xy}
\usepackage{mathrsfs}

\usepackage[framemethod=TikZ]{mdframed}

\usepackage[margin=10pt,font=small,labelfont=bf]{caption}
\usepackage{changepage}
\usepackage{setspace}
\usepackage{upgreek}

\numberwithin{equation}{section}

\pgfplotsset{compat=1.18}

\newcommand{\address}[1]{\vbox{\center\em#1}}

\usepackage{tikz}
\newcommand*{\boxcolor}{black}
\makeatletter
\renewcommand{\boxed}[1]{\textcolor{\boxcolor}{%
\tikz[baseline={([yshift=-1ex]current bounding box.center)}] \node [rectangle, minimum width=1ex,rounded corners,draw] {\normalcolor\m@th$\;\,\displaystyle#1\;\,$};}}
 \makeatother

\begin{document}

\thispagestyle{empty}


\begin{center}

{\LARGE \bf {Study of chaos and scrambling in hairy AdS Soliton}}
\end{center}

\bigskip \noindent

\bigskip

\begin{center}

Adrita Chakraborty,${}^{a}$ Balbeer Singh${}^b\,$ 

\address{
a)Faculty of Physics and Applied Computer Science, \\AGH University of Krakow, al. A. Mickiewicza 30, 30-059 Krakow, Poland\\\vspace{.3cm}
b) Department of Physics, Indian Institute of Technology Kharagpur, Kharagpur - 721302, India\\\vspace{.3cm}
}

\vspace{0.15in}
    
{\tt achakraborty@agh.edu.pl, curiosity1729@kgpian.iitkgp.ac.in
}

\bigskip

\vspace{1cm}

\end{center}

In this work, we perform a comprehensive study of the classical and quantum chaos in a candidate five-dimensional hairy AdS soliton. It is a horizonless geometry holographically dual to a confining field theory with finite scalar potential. We probe classical chaos by using particle geodesics and closed classical string. While the former shows no signature of chaos, the latter provides chaotic dynamics of the string using the Lyapunov exponent and the evolution of the Poincaré section. We perform an independent spectral analysis using the tools of the random matrix theory (RMT), namely the level space distributions and the Dyson-Mehta(DM) $\Delta_{3}$-statistics. We observe a clear transition from the low energy Wigner-Gaussian Orthogonal Ensemble (GOE) distribution to the high energy Poisson distribution. This signifies a flow from quantum chaos in the infrared to integrability in the ultraviolet. We quantitatively characterize the inherent quantum scrambling in the dual theory by computing the butterfly velocity, the rate of spatial spread of the information scrambling, inside the bulk. We undergo two independent holographic methods -- entanglement wedge reconstruction and derivation of out-of-time-ordered correlators via shockwave analysis. In these methods, we heuristically consider the region near the soliton tip to provide the infrared physics of scrambling in analogy with   the near-horizon region of a black hole. We find that the hair parameter controls various scrambling properties. Finally, we make comments on the interplay between insulator/superconductor phase transition in hairy soliton geometry and dynamical transition from integrability to chaos as both of these are affected by the presence of the hair parameter.

\newpage

\setcounter{tocdepth}{2}
{}
\tableofcontents

\newpage



\section{Introduction}\label{sec:intro}
The study of chaos is one of the most significant and efficient tools for understanding the physics of any disordered system. Observables of a chaotic system exhibit high sensitivity to very small changes in its initial conditions. In disordered classical systems, the manifestation of chaos is well-established by means of exponential separation of nearby phase-space trajectories. The Lyapunov exponent $\lambda_L$ of such system quantifies the rate of this divergence as
\begin{equation}
\lvert\delta x(t)\rvert \sim \lvert\delta x(0)\rvert e^{\lambda_L t},~~
\lambda_L = \lim_{t\to\infty}\frac{1}{t}\ln\frac{\lvert\delta x(t)\rvert}{\lvert\delta x(0)\rvert}\,.
\end{equation}The existence of positive and real $\lambda_L$ is the hallmark of sensitive dependence on initial conditions and hence chaotic dynamics.  

In quantum mechanical systems, defining chaos is more subtle, since the unitary time evolution forbids the classical notion of diverging trajectories.  The chaotic behaviour in quantum systems is instead diagnosed by using operator growth and spectral statistics. Random Matrix theory (RMT) \cite{c9eb8278-b5e2-37bc-a322-8e81785f98ed} provides a convenient tool for the spectral analysis of quantum chaos. In this framework, the eigenvalue distribution follows a level space repulsion similar to the Gaussian Orthogonal Ensemble(GOE) for chaotic system \cite{BERRY1981163,Bohigas:1988arf} whereas that of integrable system coincides with Poisson statistics \cite{Berry1977LevelCI}. Quantum chaos in a thermal system is typically understood through the growth of noncommutativity between time-evolved operators. The key diagnostic is the double commutator,  \cite{Shenker:2013pqa}
\begin{align}
    C_\beta(t) \equiv \langle \| [V(t), W(0)] \|^2 \rangle_\beta = 2 - 2 \operatorname{Re} \langle V^\dagger(t) W^\dagger(0) V(t) W(0) \rangle_\beta\,,
    \label{otoc general}
\end{align}between two generic Hermitian operators $V$ and $W$ separated by time $t > 0$. Here, $\langle.\rangle$ denotes the thermal expectation value in a canonical ensemble at the inverse temperature $\beta$, with $|O|^2 \equiv O^\dagger O$. The second term in (\ref{otoc general}) is the out-of-time-ordered correlator(OTOC) which decays exponentially in a nonchaotic system but grows as 
\begin{align}
     C_\beta(t) \sim e^{\lambda_L (t - t_*)},
\end{align}
for $t \gtrsim t_{*}$ in the chaotic regime with a bound on the quantum Lyapunov exponent $\lambda_L \leq 2\pi/\beta$ \cite{Shenker:2013pqa,Sekino:2008he, Maldacena:2015waa}. The scrambling time is denoted by $t_*$. For local operators $V(x,t)$ and $W(0,0)$ with temporal, as well as, spatial separations, the commutator spreads as\cite{Shenker:2014cwa}
\begin{align}
    C_\beta(x,t) \equiv \langle \| [V(x,t), W(0,0)] \|^2 \rangle_\beta 
\;\sim\; e^{\lambda_L \,(t - t^* - |x|/v_B)}
\end{align} where $v_{B}$ is the butterfly velocity that measures the maximal speed at which the spatial scrambling of a localized perturbation propagates. This is the velocity of the wavefront for the region where $C_{\beta}(x,t)$ becomes of order unity. Out-of-time-ordered correlators are analyzed in \cite{Hashimoto:2017oit} as a measure of quantum chaos for quantum mechanics with a generic Hamiltonian. Out-of-time-ordered correlators are studied in the context of various critical systems, e.g., 2D  CFT \cite{Khetrapal:2022dzy}, critical Ising chains \cite{PhysRevB.97.144304} driven CFT \cite{Das:2022jrr}, systems with Lifshitz criticality \cite{Jahnke:2017iwi,Plamadeala_2018}, noncommutative gauge theory \cite{Fischler:2018kwt} and so on. 
In recent years, chaos has become a central diagnostic for investigating the holographic paradigm \cite{Roberts:2014isa,Maldacena:2015waa,PandoZayas:2010xpn,Basu:2011di}. The development of the AdS/CFT correspondence \cite{Maldacena:1997re,Gubser:1998bc,Witten:1998qj} and subsequently the other gauge/gravity dualities \cite{tHooft:1999rgb,Aharony:1999ti}, establish a precise mapping between a thermal state of the boundary QFT and a black hole geometry in the bulk. The rapid thermalization of a localized boundary perturbation is holographically encoded in the fast scrambling dynamics of the black hole. Extensive research has so far been conducted on the chaotic black hole dynamics \cite{Bombelli:1991eg,Shenker:2013pqa,Shenker:2014cwa,Shenker:2013yza,Roberts:2014isa,Hashimoto:2016dfz,Grozdanov:2017ajz,Hashimoto:2018fkb,Cubrovic:2019qee,Dutta:2023yhx,Nayak:2025kbl,Jahnke:2017iwi,Ahn:2019rnq,Jahnke:2019gxr,Dutta:2024rta,Dong:2022ucb,Fischler:2018kwt,Craps:2021bmz,Karan:2023hfk,Baishya:2024gih,Lilani:2024fth,Kaushal:2025rqo,Chua:2025vig,Blake:2016wvh,Chakrabortty:2022kvq,Jeong:2024jjn,Ahn:2025exp}.\\

However, gravitational backgrounds without horizons, such as AdS solitons, present a fascinating and less-explored arena.  The AdS soliton geometry was first established by Horowitz and Myers in \cite{Horowitz:1998ha} via a double analytic continuation of near-extremal Schwarzschild–AdS black hole solutions. Solitonic geometries are smooth and horizonless geometry with a compactified spatial circle. The gravitational energy of the compactified circle is negative relative to AdS, matching the Casimir energy of dual field theory. Such solutions attain perturbative stability and serve as the ground state with anti-periodic boundary conditions of fermions around the circle. During double Wick rotation, which relates the soliton to the black hole geometry, the smoothening tip of the soliton coincides with the horizon of the black hole. While the soliton globally bears a different structure from the black hole geometry,
 the tip of the solitons acts as an effective infrared obstruction in the associated black hole configuration. Thereby, the near-tip (near-horizon) region of a soliton (black hole) consistently describes the lower energy IR region of the dual thermal field theory. For black hole, the spectrum of the dual theory remains continuous everywhere from IR to UV. However, for a soliton, the spectrum acquires finite gap in the IR region, making the dual theory confined. As soon as the radius of compactification in the soliton geometry reaches that of the black hole, the gap in the field theory spectrum vanishes in UV causing  deconfinement. Till date, various extensions of AdS soliton geometries have been proposed numerically and analytically as consistent solutions of different gravity theories \cite{Dias:2011tj,Brihaye:2013tra,Ogawa:2011fw,Shi:2016bxz,Cadoni:2011yj,Brihaye:2012ww,Smolic:2015txa,Kleihaus:2013tba,Anabalon:2016izw,Anabalon:2021tua,Canfora:2021nca,Anabalon:2022aig,Anabalon:2024che,Anabalon:2022ksf}. Gravity backgrounds with soliton structures provide a natural holographic laboratory for confinement, mass gaps, and other related spectral phenomena \cite{Aharony:1999ti,Polchinski:2000uf,Witten:1998zw}. In such a holographic framework, the bound states of gluons, often known as glueballs, manifest themselves as the discrete quantum spectrum of bulk fluctuations at large N and strong 't Hooft coupling limits. Several works \cite{Pullirsch:1998wp,Markum:1997fk,Basu:2013uva, Basu:2011dg, Zhou:2018rac, Akutagawa:2018yoe,Natsuume:2023hsz,Natsuume:2023lzy,Giataganas:2017guj,Giataganas:2017koz,Baishya:2023ojl,Shukla:2023pbp,Shukla:2024wsu,Lilani:2025wnd} have investigated the chaotic properties of confining backgrounds, including different solitonic geometries and different QCD-like theories. Most of these studies demonstrate signatures of chaos at the classical level. More recently, non-perturbative studies of pole-skipping in AdS solitons analyzed the master equations governing gravitational perturbations \cite{Natsuume:2023hsz,Natsuume:2023lzy}, revealing the appearance of so-called ``missing states". However, because of the absence of a proper horizon, a suitable interpretation of quantum chaos in these classes of backgrounds remains blurred.
Very recently, quantum complexity was also explored for AdS soliton bulk, where late-time growth of complexity was found to be oscillatory \cite{Anegawa:2024wov}.\\ 

Our present work focuses on a specific class of analytical hairy AdS soliton in five dimensions \cite{Anabalon:2016izw}. The hair sourced by a non-trivial dilaton field introduces an additional parameter $\alpha$ that consistently deforms the geometry, offering a rich playground to explore the interplay between confinement, phase transitions, and the emergence of chaos.
Therefore, a central motivation behind our study is to understand how the scalar hair influences the chaotic dynamics in such a background. Then we examine the phase transition through chaos and information scrambling. To this end, we employ a heuristic analogy, treating the smooth tip of the soliton geometry as an effective horizon, motivated by the double Wick rotation, which relates it to a black hole. Although the soliton is globally horizonless, this analogy provides a practical tool to probe scrambling in its deep infrared region.
 Classically, we probe the chosen geometry with both particles and closed strings. Particle motion remains integrable due to the suppression of the chaos in the absence of a horizon. At this stage, we start to treat the tip of the AdS soliton analogously to the event horizon of its cousin black hole. It is trustworthy to consider the region near the tip to study the information scrambling in the IR sector of the dual thermal field theory. Such an assumption gives rise to a stark contrast while using a classical closed string as the probe. The extended nature of strings reveals clear chaotic behaviour in Poincar\'e sections, especially at higher energies.  
 Speaking of quantum diagnostics, a well-known spectral signature of chaos arises from the tools of  random matrix theory. We choose spectral statistics specifically, the level spacing distribution and the Dyson-Mehta $\Delta_3$ statistics \cite{Lozej:2021jgt,Balasubramanian:2024ghv} for the spectrum of a probe string quantized via mini-superspace quantization scheme\cite{PhysRevD.28.2960,Seiberg:1990eb,Douglas:2003up}. Both of these techniques identify a clear emergence of a dynamical transition, from chaos, designated by the Wigner-GOE distribution, to orderedness, designated by Poisson distribution, of the eigenvalue spectrum from lower to higher energies. 
These statistical methodologies eventually complement the analytical holographic methods of understanding quantum scrambling in boundary information. In our article, we compute the butterfly velocity using two complementary holographic techniques: entanglement wedge (EW) reconstruction and derivation of OTOC via shockwave analysis on a double-sided extension of the hairy AdS soliton geometry. In these methods,  the quantity $\lambda_L = 2\pi / \beta$ should not be interpreted as the Lyapunov exponent originating from a Hawking temperature but rather as a coefficient containing the compactification $\beta$ of the soliton and characterizing the information scrambling suitably in such system. 
We mention necessary remarks on these quantities while interpreting the dual field theoretic quantities in the subsequent sections. 
The EW method reveals how bulk perturbations disrupt boundary entanglement, whereas the shockwave analysis directly computes the exponential spreading of perturbations. From the dual perspective, such spreading indicates toward the scrambling of information in the boundary data in the dual gapped field theory which has certain thermal properties. Both of these holography-based methods yield positive and real butterfly velocity $v_B$.  
 We numerically observe the behaviour of the non-zero butterfly velocity $v_{B}$ of the scrambling in our desired background with the strength of the hairy effect. Our final comments shed light into a possible interplay between different phase transitions in hairy AdS soliton background and a dynamical transition from integrability to chaos emerging due to the presence of the scalar hair.\\ 

The structure of the paper is as follows. In Section \ref{brief-hairy-ads-soliton}, we outline the construction of the hairy AdS soliton and highlight its key properties. In Section \ref{sec:method}, we investigate classical chaos using two distinct probes: particle geodesics and classical closed strings. These two analyses yield sharply contrasting outcomes: the particle geodesic shows no signs of classical chaos, while the classical closed string clearly exhibits chaotic behaviour. In Section \ref{numerical-apprach}, we adopt a numerical approach to quantum chaos by studying the hairy AdS soliton within the mini-superspace quantization framework. Here, we employ two standard spectral diagnostics of quantum chaos: the level spacing distribution and the Dyson–Mehta statistics. In Section \ref{sec:quantumchaos}, we focus on the study of quantum scrambling, where we compute the butterfly velocity using two complementary approaches, shockwave analysis and entanglement wedge reconstruction. In Section \ref{sec:critical phase transition}, we make some comments on the critical phase transition through the lens of chaos and discuss its implications for the corresponding scrambling, including the dynamical flow between chaos and integrability. Finally, we conclude with a summary and discussion in Section \ref{conclusion}.
\section{Review: 5D Hairy AdS soliton} \label{brief-hairy-ads-soliton}
In this section we present a brief review on hairy AdS soliton solutions in Einstein-dilaton gravity theory. It is well established through a plethora of research that there exist AdS soliton geometry which are derived by incorporating double Wick rotation along time and one of spatial directions in a planar AdS black hole. Such solutions are regular and asymptotically merge to planar AdS spacetime. Hairy AdS solitons can be similarly derived from hairy AdS black hole geometry. An analytically consistent candidate 5D metric for such black hole geometry can be considered as \cite{Anabalon:2016izw} 
\begin{subequations}
\begin{align}
    &ds^2 = \Omega(x)\left[-f(x)dt^2+\frac{\eta^2 dx^2}{f(x)}+\frac{dx_1^2}{l^2}+\frac{dx_2^2}{l^2}+\frac{dx_3^2}{l^2}\right]\\&
    \Omega(x) = \frac{25 x^4}{\eta^2(x^5-1)^2},~~f(x)=\frac{1}{l^2}+\frac{\alpha}{3^2 10^4}\left(x^{10}-6 x^5+30 \log x+3+\frac{2}{x^5}\right)
    \end{align}
    \label{hairy black hole 5d}
\end{subequations}The conformal boundary of the above geometry is at $x=1$. 
 This is a legitimate solution of the Einstein-Maxwell-Dilaton gravity action 
\begin{equation}
    I=\int_{\mathcal{M}}d^5x\sqrt{-g}\left[R-\frac{1}{2}\partial_{\mu}\phi\partial^{\mu}\phi-V(\phi)\right]-\frac{1}{\kappa}\int_{\partial\mathcal{M}}d^4x\sqrt{-\gamma}K
    \label{bulk action}
\end{equation}where, $V(\phi)$ is a scalar potential that causes the appearance of hair in the corresponding background solution and $\kappa=8\pi G_N$. $K$ defines the extrinsic curvature of the boundary of the manifold $\mathcal{M}$ and $\gamma$ is the determinant of the induced metric on the boundary. The explicit functional form of the potential $V(\phi)$ is given in \cite{Acena:2012mr,Anabalon:2016izw} which contains the intrinsic hairy parameter $\alpha$. For $x>0$, the stability of the solution (\ref{hairy black hole 5d}) requires only negative values of $\alpha$. Note that, as the hairy parameter $\alpha$ vanishes, the above geometry asymptotically merges to conformal geometry 
with curvature scalar $R=-\frac{4 \left(33 x^{10}+84 x^5+8\right)}{25 l^2 x^6}$. The conformal factor $\Omega(x)$ associated with the metric (\ref{hairy black hole 5d}) is specified when integrating the equations of motion of the action (\ref{bulk action}). We particularly choose the conformal factor $\Omega(x)$ and $f(x)$ for the metric as what was considered in \cite{Anabalon:2016izw} for $\nu=5$. The exact form of the conformal factor helps designating a scalar potential that has a logarithmic dependence on the radial coordinate in the bulk gravity theory. The geometry of hairy AdS soliton is obtained by applying a double Wick rotation such as $t\rightarrow i\theta$ and $x_1\rightarrow i\tau$ so that it reduces to \cite{Horowitz:1998ha} 
\begin{equation}
    ds^2 = \Omega(x)\left[-\frac{d\tau^2}{l^2}+\frac{\lambda^2 dx^2}{f(x)}+f(x) d\theta^2+\frac{dx_2^2}{l^2}+\frac{dx_3^2}{l^2}\right], 
    \label{hairy AdS soliton 5d}
\end{equation}
where the integration constant is designated as $\lambda$ to make it different from $\eta$ of the hairy AdS black hole. The conformal factor $\Omega(x)$ takes the form of
\begin{equation}
    \Omega(x) = \frac{25 x^4}{\lambda^2(x^5-1)^2}
    \label{conf factor for hairy soliton}
\end{equation}Nevertheless, the blackening factor $f(x)$ remains the same as it is fully a function of the radial scale $x$ of the hairy AdS black hole geometry, and there is no Wick rotation applied along the radial direction. Due to Wick rotations along the temporal and one of the spatial directions, the geometry assumes a cigar-like structure without any horizon and is capped off at $x=x_s$ along the radial scale, where
\begin{equation}
    f(x)|_{x=x_s}=\frac{1}{l^2}+\frac{\alpha}{3^2 10^4}\left(x_s^{10}-6 x_s^5+30 \log x_s+3+\frac{2}{x_s^5}\right)=0
\end{equation}In case of a hairy black hole, it is the same point on the radial scale where the blackening factor vanishes and the horizon $x_h$ of the black hole appears. Hence, the resulting bulk geometry extends in the range, $1<x\leq x_s$, along the positive radial direction. The cigar-like structure causes an emergent IR cutoff and eventually a dynamical energy scale in the corresponding theory \cite{Witten:1998zw,Aharony:1999ti}. As all the spatial directions and the time direction acquire the same dimension in the hairy black hole geometry, the analytic continuation with the double Wick rotation does not affect the flat structure of the corresponding boundary geometry. Although there is no intrinsic interpretation of temperature in soliton geometries due to the absence of a horizon, the notion of temperature as well as thermodynamics of such a system appears from the compactification along the Euclidean time direction while applying double analytic continuation, the circumference of compactification being $\beta$, which is chosen suitably to avoid the conical singularities of such geometry. This $\beta$ is interpreted as the inverse of temperature and plays a significant role in the thermodynamics and phase transition in such geometries \cite{Horowitz:1998ha}.
For hairy AdS soliton background given in (\ref{hairy AdS soliton 5d}), the radius of compactification $\beta$ and hence the associated temperature $T$ assumes nontrivial dependency on $\alpha$. Solitonic geometries find their own significance in studying confining properties in the dual boundary field theory as there appears certain mass gap in the infrared regime due to its cigar-like construction. Thus they provide a promising stage to understand the confinement/deconfinement-like critical phase transition under the IR to UV flow. Added to this, there occurs a distinct second order phase transition between holographic insulating to superconducting phases once a scalar hair is introduced in a soliton-like geometry \cite{Peng:2011gh,Peng:2016mxd,OuYang:2020mpa}. In the following sections, we will study the emergence of scrambling in the dynamics of our desired soliton geometry using various seemingly different approaches. We will subsequently understand the fate of the scrambling and its effect on various phase transitions. 

   %



\section{Classical chaos}\label{sec:method}
In this section, we will present the notion of chaos in a hairy AdS soliton background from classical perspective, more precisely, using the motion of a probe free falling particle and a probe fundamental string.
\subsection{Particle motion and chaos} 
 The motion of a free-falling particle in an AdS soliton has been understood in \cite{Zhou:2018rac} where it is shown that, the chaos in such particle dynamics is suppressed in such a geometry without a horizon. Here, we will study the motion of such free particle in a hairy AdS soliton background. For a free particle with mass $m$ and zero potential $(V = 0)$, the action is:
\begin{align}
    S_p = -m \int \sqrt{-g_{\mu\nu} \dot{X}^\mu \dot{X}^\nu} \, dt,
\end{align}
where $\dot{X}^\mu = \frac{dX^\mu}{dt}$ is the velocity with respect to the parameter $t$. We use the static gauge $\tau = t$, so $\dot{\tau} = 1$, and assume radial motion only, $x=x(\tau)$, the rest of the coordinates are constant: $\theta=x_{2}=x_{3}=\text{constant}.$\\
The Lagrangian is:
\begin{align}
    L = -m \sqrt{\Omega(x) \left( \frac{1}{l^2} - \frac{\lambda^2 \dot{x}^2}{f(x)} \right)}.
\end{align}
Since $L$ is independent of $t$, the energy $E$ is conserved 
\begin{align}\label{energy}
    E&= \frac{\partial L}{\partial \dot{t}} 
    = m \frac{\sqrt{\Omega(x)} \frac{1}{l^2}}{\sqrt{\frac{1}{l^2} - \frac{\lambda^2 \dot{x}^2}{f(x)}}}.
\end{align}
and canonical momentum:
\begin{align}
    p_x &= \frac{\partial L}{\partial \dot{x}}
    =m \frac{\sqrt{\Omega(x)} \frac{\lambda^2 \dot{x}}{f(x)}}{\sqrt{ \left( \frac{1}{l^2} - \frac{\lambda^2 \dot{x}^2}{f(x)} \right)}}.
\end{align}
From eq \ref{energy}, solve for $\dot{x}$, we get
\begin{align}
    \frac{dx}{dt} = \sqrt{\frac{f(x)}{\lambda^2} \left( \frac{1}{l^2} - \frac{\Omega(x)}{A^2} \right)}.
    \label{particle velocity}
\end{align}
where we define $A^2 = l^4 \left( \frac{E}{m} \right)^2$. We define the Rindler momentum as 
\begin{align}
    p_\rho = mA \sqrt{\frac{1}{l^2} - \frac{\Omega(x)}{A^2}},
\end{align}
It is important to note that the expression $  \Big(\frac{1}{l^2} - \frac{\Omega(x)}{A^2}\Big)$ under the square root must be positive for a real value of $\dot{x}$. Therefore, it restricts the parameter space.
\begin{figure}[h!]
    \centering
    \includegraphics[width=0.5\linewidth]{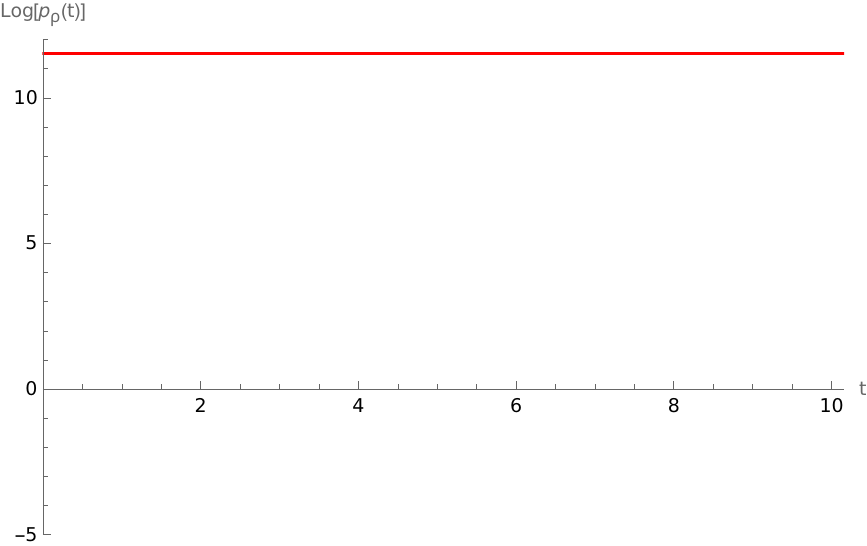}
    \caption{Time dependence of the Rindler momentum $p_{\rho}(t)$ at the parameters $ \alpha=-0.1,\, x_{s}\approx 2.50,\, \lambda= 1,\,  x_{0}\approx 2.30$. We set $l=10,\, m=1,\, E=5$.}
    \label{prho-plot}
\end{figure}Some observations are straightforward from the above analysis. To begin with, it becomes clear that the tip of the geometry is solely determined by parameters $l \text{ and } \alpha$. We will explore more on this while studying quantum chaos via both analytic and numerical approaches later on. Fixing the parameters $l,\alpha$, $\lambda$ and  $m$, the Rindler momentum $p_{\rho}$ can either become real or complex depending on the values of $ E$ and $ x_{0}$ that set the initial conditions on the ordinary differential equation. If sufficiently high energy is provided, one may achieve an allowed region of real values of the momentum; otherwise, the momentum becomes complex. This is true for both the near-tip region and the far-from-tip region. Also, we can see that $\frac{dx}{dt} > 0$ as  $f(x) > 0$ for  $x < x_s$, and the velocity of the particle is real in (\ref{particle velocity}). It explains that the particle moves from $x_0 < x_s $ toward the tip  $x_s$. As it approaches  $x_s$,  $f(x) \to 0,$ slowing the motion, but the momentum remains well-defined and real.
However, due to the absence of the horizon in the hairy AdS-soliton geometry the chaos is suppressed as evident from the Fig \ref{prho-plot}. This can be attributed to the fact that the dynamical contribution from  $\Omega(x)$ (or the radial motion) is very negligible and so the chaos gets suppressed quite fast. For a comparative understanding of the chaotic dynamics of the hairy AdS solitonic system, we will probe the background with a fundamental string in the next section.

\subsection{Chaos from stringy perspective}\label{sec:classicalchaos}
\begin{figure}[h]
    \centering
    \includegraphics[width=0.42\linewidth]{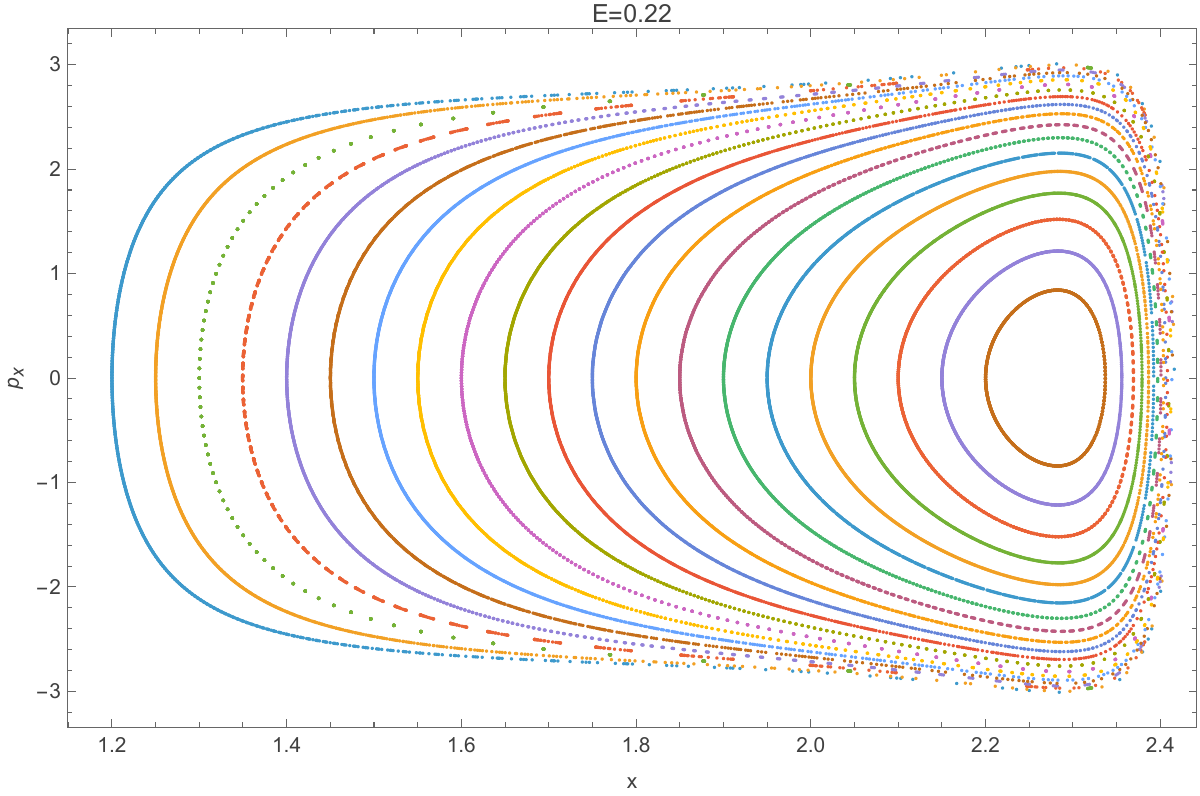}
    \includegraphics[width=0.42\linewidth]{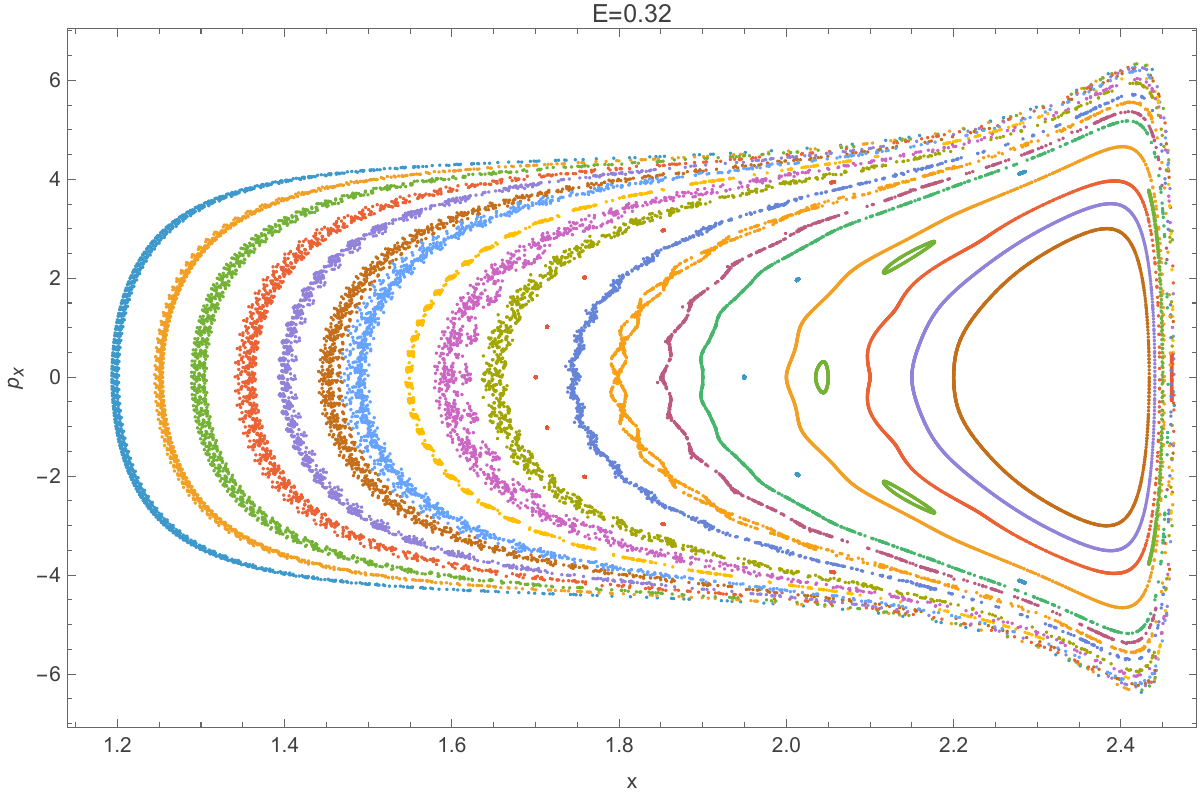}
    \includegraphics[width=0.42\linewidth]{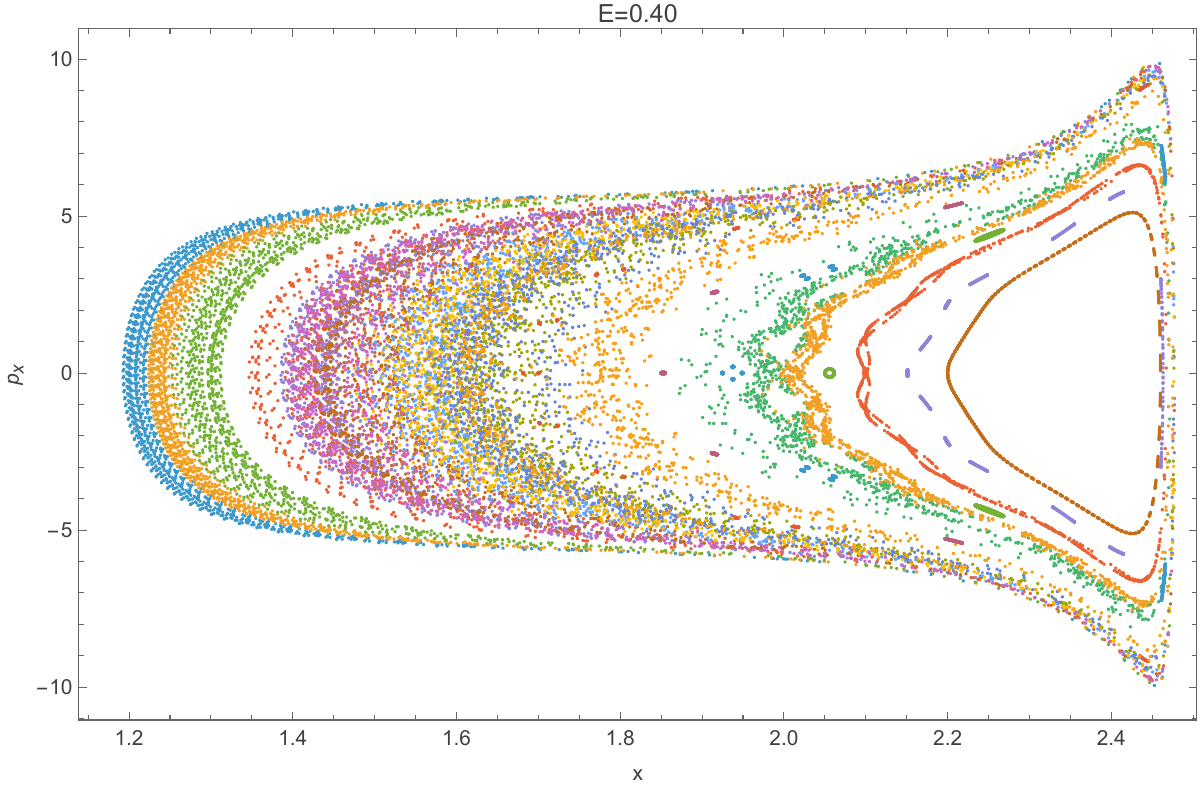}
    \includegraphics[width=0.42\linewidth]{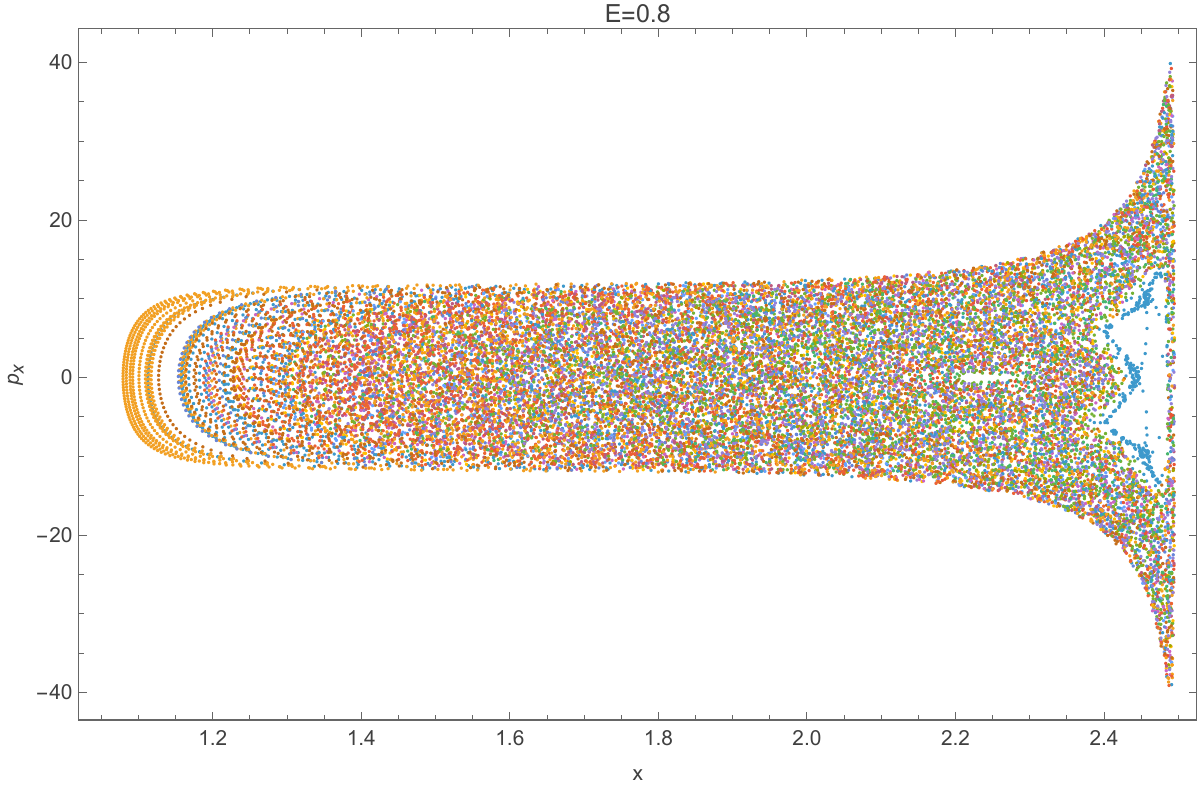}
    \includegraphics[width=0.42\linewidth]{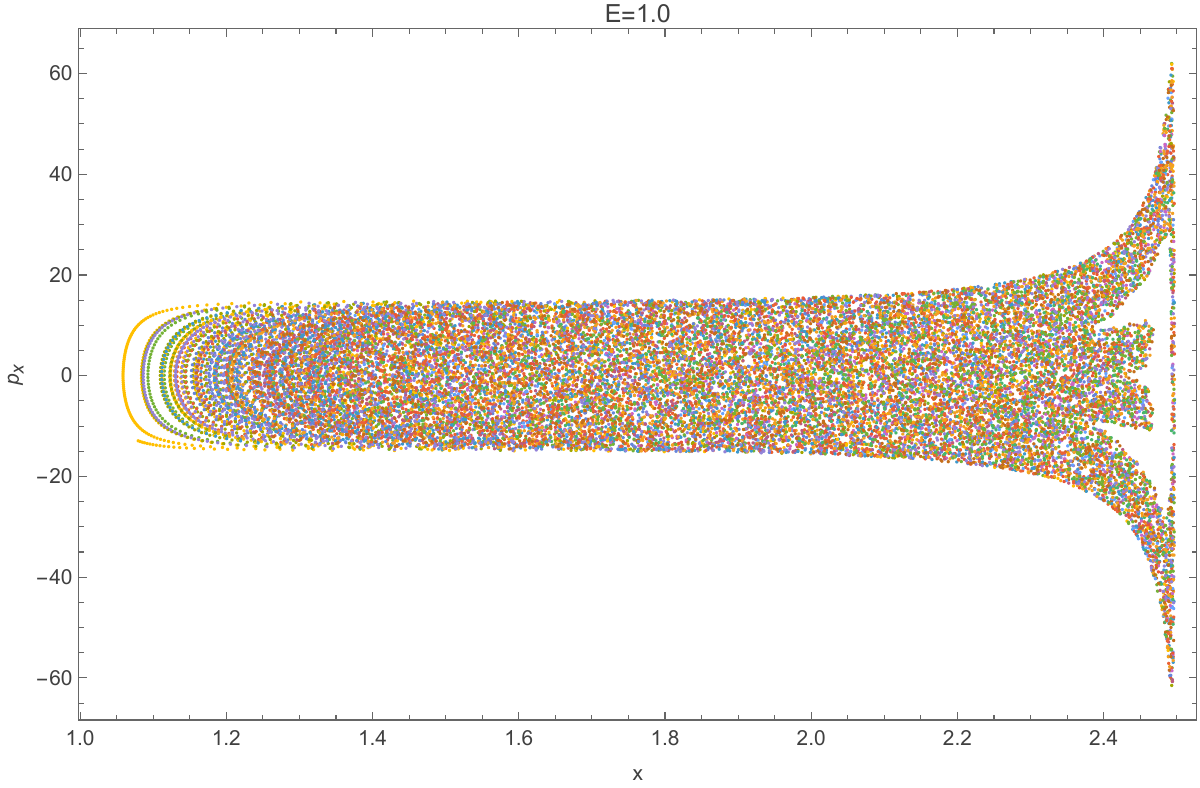}
    \includegraphics[width=0.42\linewidth]{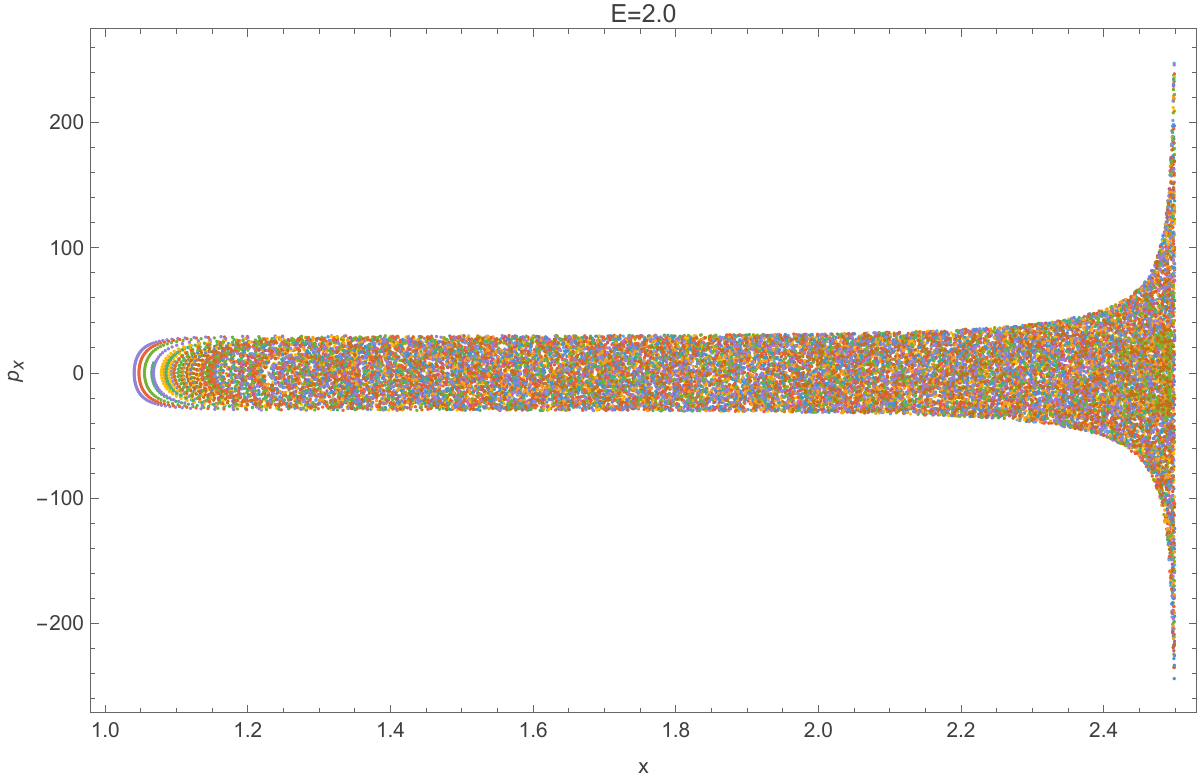}
    \caption{Poincar\'e sections for hairy AdS soliton for different energies at parameters: $n=1\,,\, l=10\,,\, \lambda=0.15\,,\, k=0.12\,,\,\alpha=-0.1$ with inital conditions $x(0)=1.2$\,,\,$s(0)=0$\,,\,$p_{x}(0)=0$. }
    \label{fig:hairy-ads-soliton-poincare}
\end{figure}

In this section, we refine our analysis of particle chaos by drawing an analogy between our geometry and the hairy AdS black hole, treating the tip of the soliton as the analogue of a black hole horizon and examine the dynamics of a closed string hovering over the hairy $AdS_{5}$ soliton background geometry. The Polyakov action for the closed string governing the dynamical evolution is given by
\begin{equation} \label{eqn: polyakov}
 S = -\frac{1}{2\pi \alpha^{\prime}} \int d\sigma d\tau \Big( \sqrt{-g}  g^{\alpha \beta} G_{\mu \nu}\partial_{\alpha} X^{\mu}\partial_{\beta} X^{\nu} - \epsilon^{\alpha \beta} \partial_{\alpha} X^{\mu} \partial_{\beta} X^{\nu} B_{\mu \nu} \Big),
 \end{equation}
 where $\alpha' = l_s^{2}$, and $l_s$ denotes the fundamental string length. The variables $X^{\mu}$ represent the embedding functions of the string worldsheet into the target space, $G_{\mu\nu}$ corresponds to the target-space metric, and $g_{\alpha\beta}$ denotes the intrinsic worldsheet metric. By exploiting the reparameterization and Weyl invariances of the action, one may consistently impose the conformal gauge, $g^{\alpha\beta} = \eta^{\alpha\beta}$. In this gauge, the requirement that the worldsheet energy--momentum tensor vanishes, $T_{\alpha\beta}=0$, leads to the Virasoro constraints:
\begin{flalign}
G_{\mu\nu}\,\partial_{\tau}X^{\mu}\,\partial_{\sigma}X^{\nu} &= 0, 
\label{eqn:gauge_conformal}\\
G_{\mu\nu}\Big(\partial_{\tau}X^{\mu}\,\partial_{\tau}X^{\nu}
+ \partial_{\sigma}X^{\mu}\,\partial_{\sigma}X^{\nu}\Big) &= 0.
\end{flalign}
We use the following ansatz for the probe closed string
\begin{equation}
\begin{aligned}
t=t(\tau)\,, \qquad \theta=\theta(\tau)\,, \qquad x=x(\tau)\, \\
x_2=s(\tau)\cos{\phi(\sigma)}\,,\qquad x_3=s(\tau)\sin{\phi(\sigma)}\,
\end{aligned}
\label{string ansatz}
\end{equation}
where $\phi(\sigma)=n \sigma$, where $n \in\mathbb{Z}$ is the winding number enclosing the string. 
Then next we find the Hamiltonian of the system 
\begin{align}
    H= \frac{k^2 \lambda ^2 \left(x^5-1\right)^2}{100 x^4 f(x)}+\frac{\left(x^5-1\right)^2 f(x) p_x^2}{100 x^4}+\frac{25 n^2 s^2 x^4}{\lambda ^2 l^2 \left(x^5-1\right)^2}-\frac{\lambda ^2 l^2 \left(x^5-1\right)^2 \left(E^2-p_s^2\right)}{100 x^4}
\end{align}
where $E$ is the energy of the system, $k \equiv p_{\phi}$ (fixed) and $p_{s}$ are the angular momenta along the $\phi$ and $s$ directions, respectively. 
Poincaré sections serve as a reliable tool to visualize the qualitative aspects of chaotic dynamics. Although they do not allow for a quantitative measure of chaos in string dynamics, they provide valuable insights into the regions where the chaotic behaviour of the string begins to manifest.
For the Poincar\'e sections, we choose, without loss of generality, the following parametric values 
\begin{equation}
    \begin{aligned}
        \alpha'= \frac{1}{2 \pi}\,,\qquad n=1\,,\qquad l=10\,,\qquad \lambda=0.15\,,\qquad k=0.12\,,\qquad \alpha=-0.1
    \end{aligned}
\end{equation}
with the initial conditions:
        $s(0)=0, \,\,\, p_x(0)=0$.
By varying the initial position $x(0)$, the momentum $p_s(0)$ is obtained from the Hamiltonian constraint $H=0$. Figure~\ref{fig:hairy-ads-soliton-poincare} illustrates how, with increasing energy, the Poincar\'e section progressively deforms into a Cantor–like dust, a characteristic signature of chaotic dynamics. Here we set the initial position as $x(0) = 1.2$. It is evident from the Poincar\'e sections that the dynamics become more chaotic for string states at higher energies, starting from $E=0.22$ to $E=2.0$. \\ 

\begin{figure}[htbp!]
	\centering
     \includegraphics[width=0.42\linewidth]{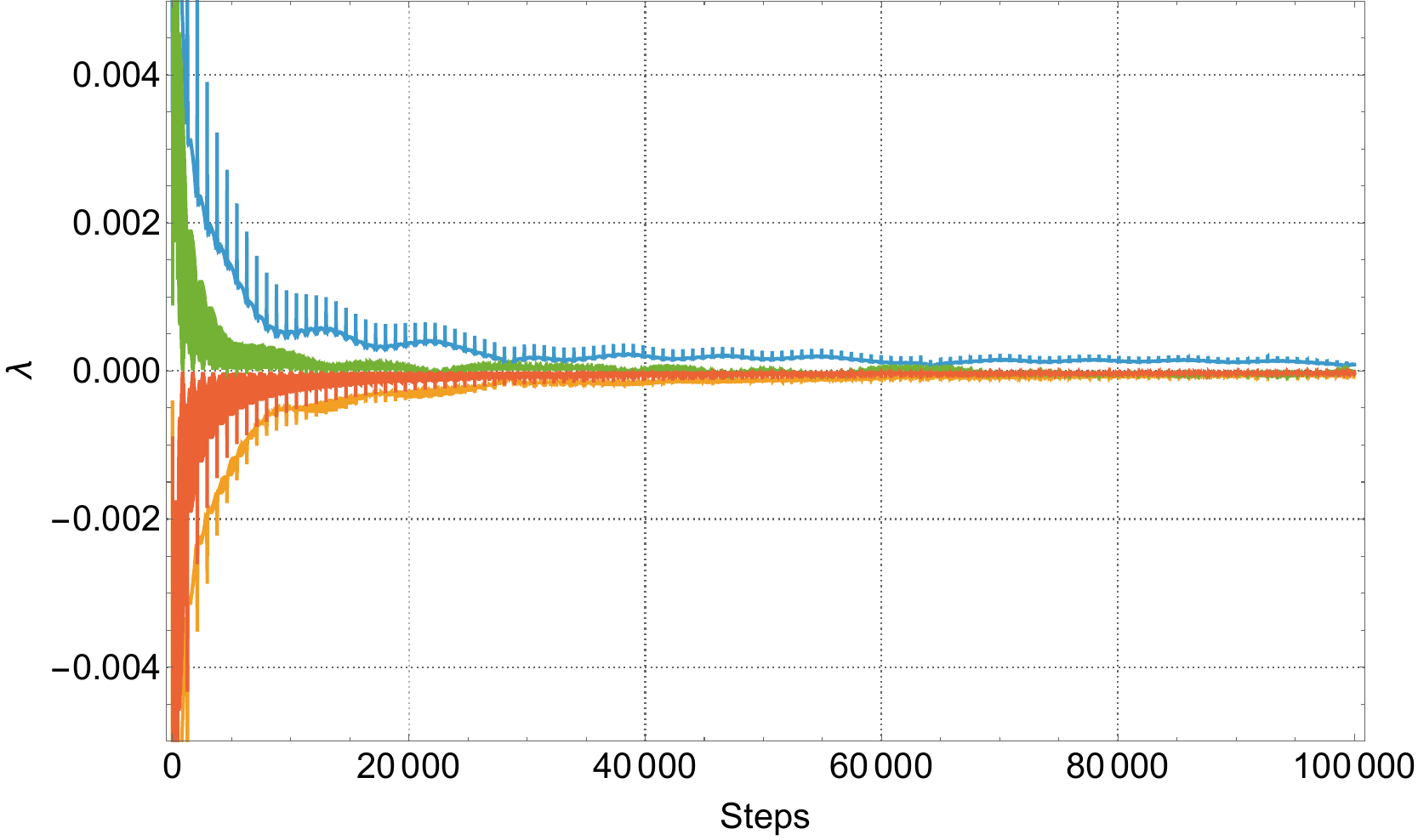}
    \includegraphics[width=0.42\linewidth]{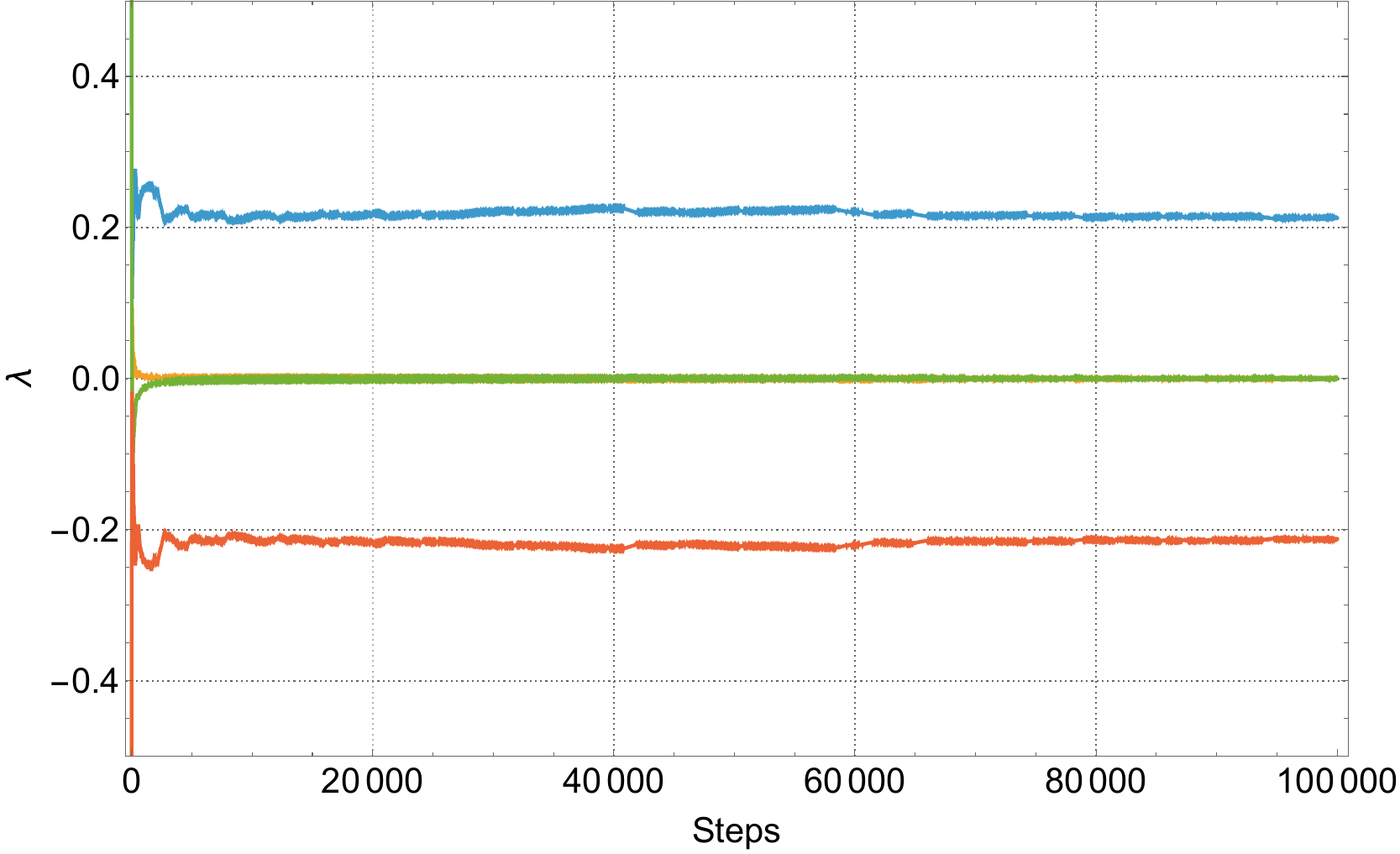}
	\caption{\label{fig:hairy-ads-soliton-lyapunov}Lyapunov exponents at low and high energy values $E=0.22$ (left) and $E=2.0$(right), respectively. The other parameter values are the same as those used in Fig \ref{fig:hairy-ads-soliton-poincare}. }
\end{figure}

%
In addition, we compute the (maximum) Lyapunov exponent $\lambda$ and check the sign of $\lambda$. Our analysis shows that the maximum Lyapunov exponent remains small in the low-energy regime, but attains a significantly positive value as the energy increases, see Fig \ref{fig:hairy-ads-soliton-lyapunov}. In our analysis, we have taken $10^5$ steps with the step size of $0.001$. At higher energy value, we observe stronger effect of chaos in the system which is consistent with the observation from the Poincar\'e sections.\\

\par It is evident from both of these analyses that the motion of a closed string probing the hairy AdS soliton background assumes more chaotic behaviour as the energy of the string states increases. Higher energy spectrum of strings appears when the string moves enough inside the bulk geometry, i.e., far enough from the boundary. 
For a solitonic background, such a region can be taken as the near-the-tip region. The generic classical string ansatz as taken above helps in constructing the 2D non-linear sigma model with the string probing the target (sub)space $AdS_{5}\times S^5$. The corresponding classical spectrum involves the functional dependency on the various parameters of the string ansatz considered. On quantization, these dispersion relations relate the anomalous dimensions of the operators in the strongly coupled nonperturbative IR sector of the boundary field theory. 
The legitimate dual theories of solitonic backgrounds are confining theories with a discrete mass gap in the associated spectrum of the bound states of gluons, often known as glueballs \cite{Aharony:1999ti,Witten:1998zw}. Therefore, the chaotic dynamics of higher energy string sprectrum implies a possible scrambling of the massive glueball operators with strong confinement in the low energy IR sector of the dual confining theory. 
 
\section{Quantum chaos via statistical approach}\label{numerical-apprach}
In this section, we present a spectral analysis of the probe string following a similar ansatz as (\ref{string ansatz}) inside the hairy AdS soliton background under consideration. The anstaz is typically chosen as  
\begin{equation}
\begin{aligned}
t=t(\tau)\,, \qquad \theta=\theta(\tau)\,, \qquad x=x(\tau)\, \\
x_2=R(\tau)\cos{\phi(\sigma)}\,,\qquad x_3=R(\tau)\sin{\phi(\sigma)}\,
\end{aligned}
\end{equation}
where $\phi(\sigma)=n \sigma$. 
The Lagrangian corresponding to the ansatz chosen is 
\begin{align}
    \mathcal{L}\sim \frac{1}{2} \Big[-\frac{\Omega(x)}{l^2} \dot t^2 + \Omega(x) f(x) \dot \theta^2 + \lambda^2 \frac{\Omega(x)}{f(x)}\dot x^2 + \frac{\Omega(x)}{l^2}\dot R^2 - \frac{\Omega(x)}{l^2}R^2 n^2\Big]
\end{align}
The conserved quantities are given by
\begin{align}
    p_t = -\frac{\Omega(x)}{l^2}\,\dot t \;\equiv\; E, 
    \;\;\;\;\;\;\;
p_\theta = \Omega(x) f(x)\,\dot\theta \;\equiv\; k
\end{align}
and,
\begin{align}
    p_R = \frac{\Omega(x)}{l^2}\,\dot R, 
\;\;\;\;\;\;\;
p_x = \frac{\lambda^2 \Omega(x)}{f(x)}\,\dot x.
\end{align}
Therefore, the canonical Hamiltonian is given by
\begin{align}
    \mathcal{H}= \frac{1}{2} \Big[-\frac{l^2}{\Omega(x)} E^2 + \frac{k^2}{\Omega(x) f(x)} + \frac{l^2 p_{R}^2}{\Omega(x)}+ \frac{f(x)p_{x}^2}{\lambda^2 \Omega(x)} + \frac{\Omega(x)}{l^2}R^2 n^2\Big]
\end{align}\subsection{Minisuperspace quantisation}
\begin{figure}[h]
     \centering
    \includegraphics[width=0.45\linewidth]{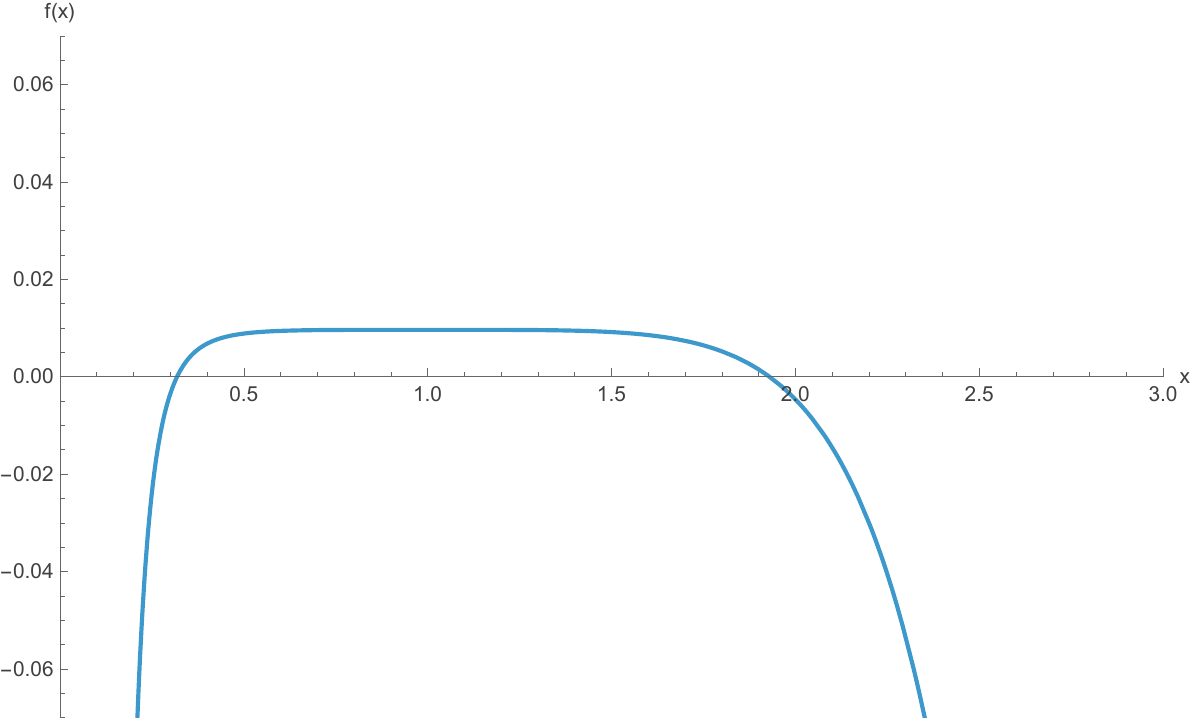}
\caption{Typically $f(x)=0$  admits two roots: one lying at small $x$ and the other at large $x$ at parameter values $\alpha=-1.5,\;\;l=10.2, \;\;  \lambda=0.1$.}
\label{roots}
\end{figure}

Now, we will quantize the spectrum of such probe string using a scheme called minisuperspaced quantization \cite{PhysRevD.28.2960,Seiberg:1990eb,Douglas:2003up} that considers only the modes associated with the centre of mass. Previously such prescription has been recently employed in \cite{PandoZayas:2012ig,Saremi:2012ji,Basu:2013uva,Shukla:2024wsu} to investigate hadronic spectra in soliton-type confining backgrounds.
The minisuperspace quantization requires a Schrodinger-like equation
\begin{equation}
    -\Delta\psi+V(x)\psi=0
\end{equation}to obtain the full quantum spectrum. The Laplacian in this equation is constructed using the metric components in the Polyakov action of the probe string. $V(x)$ denotes the potential appearing in the corresponding Hamiltonian. We start with promoting the generalized momenta to the operators as
\begin{align}
    p_{R}^2 \xrightarrow[]{} -\nabla^2_{R}, \;\;\;\;\;\;\;\; p_{x}^2 \xrightarrow[]{} -\nabla^2_{x}
\end{align}The Laplacian is calculated with respect to the effective metric components $-g_{tt}=g_{RR}= \frac{\Omega(x)}{l^2},~~
g_{\theta\theta}= \Omega(x)f(x)$ and
$g_{xx}= \frac{\lambda^2\Omega(x)}{f(x)}$ as seen in the Lagrangian.
Thus, the minisuperspace Hamiltonian takes the form as
\begin{align}
    \mathcal{H}= \frac{1}{2}\Big[-\frac{E^2 l^2}{\Omega(x)} + \frac{k^2}{\Omega(x) f(x)} - \frac{1}{2}\frac{1}{\lambda^2 \Omega(x)}\frac{df(x)}{dx}\partial_{x}- \frac{f(x)}{\lambda^2 \Omega(x)}\partial_{x}^2-\frac{l^2}{\Omega(x)}\partial_{R}^2 + \frac{\Omega(x)R^2 n^2}{l^2} \Big]
\end{align}
The eigenvalue equation $\mathcal{H}\psi=0$ then becomes,
\begin{align}
    E^2 \psi(x,R) =-\partial_{R}^2\psi(x,R) - F(x)\partial_{x}^2\psi(x,R) - G(x) \partial_{x} \psi(x,R) + V_\text{eff}(x,R)\psi(x,R)
\end{align}
where, 
\begin{align}
     F(x)\equiv \frac{f(x)}{\lambda^2 l^2},\;\;\;\;\;\;  G(x)\equiv \frac{1}{2\lambda^{2} l^2}\frac{df(x)}{dx}, \;\;\;\;\;\; 
    V_{\text{eff}}(x,R) \equiv \frac{k^2}{l^2 f(x)} + \frac{\Omega(x)^2 R^2 n^2}{l^4}
\end{align}
With a coordinate transformation $dy=\frac{dx}{\sqrt{F(x)}}$. Then one gets
\begin{align}
    E^2 \psi = -\partial_R^2 \psi + \left( \frac{F'(x)}{2 \sqrt{F(x)}} - \frac{G(x)}{\sqrt{F(x)}} \right) \partial_y \psi - \partial_y^2 \psi + V_{\text{eff}}(x, R) \psi
\end{align}
which simplifies to
\begin{align}
    E^2 \psi(y, R) = -\partial_R^2 \psi(y, R) - \partial_y^2 \psi(y, R) + V_{\text{eff}}(x(y), R) \psi(y, R),
    \label{eq-eigen value eq}
\end{align}
where $x(y)$ is the inverse of $y(x)$, and:
\begin{align}
    V_{\text{eff}}(x(y), R) = \frac{k^2}{l^2 f(x(y))} + \frac{\Omega(x(y))^2}{l^4} n^2 R^2.
\end{align}
Now, to find the range of the $x$, if we look at the function $f(x)$, two roots are generally possible, as shown in Fig \ref{roots}, depending on the parameter values up to certain critical values of the $\alpha$ and $l$. 
Thus, the plausible range of $x$ is $[x_{s}, x_{s\;max}]$ where $f(x)$ is non-negative. 
Therefore, the range of $y$ becomes $[0,y_{max}]$. For instance, for an arbitrary parameter choice: $\alpha=-1.5\;\;l=10.2, \;\;  \lambda=0.1$, the range of y is $[0, 16.801]$. In the next subsection, we use the level space distribution and the Dyson-Mehta statistics as the quantum chaotic indicators to check the chaos in the system. \\
\subsection{Level space distribution and Dyson-Mehta statistics}
The level spacing distribution is primarily concerned with short–range correlations between consecutive eigenvalues. After unfolding the spectrum so that the average spacing is normalized, one examines the probability distribution $P(s)$ of nearest–neighbor spacings $s$. In systems that are integrable, the levels are essentially uncorrelated and follow a Poisson distribution, 
\begin{equation}
    P(s)=e^{-s},
\end{equation}
characterized by the absence of level repulsion. Conversely, in quantum chaotic systems, the spacings obey Wigner–Dyson statistics, with the GOE being typical for time-reversal symmetric systems. In this case, the distribution takes the Wigner surmise form 
\begin{equation}
    P(s)=\frac{\pi}{2}s e^{-\pi s^{2}/4},
\end{equation}
exhibiting a clear level repulsion at small spacings.\\

The Dyson–Mehta $\Delta_{3}(L)$ statistic, on the other hand, measures long–range spectral correlations by quantifying the deviation of the cumulative spectral density from a best–fit straight line over an interval of length $L$. In the case of a random spectrum following Poisson statistics, the spectral rigidity is 
\begin{equation}
    \Delta_{3}(L) = \frac{L}{15}.
\end{equation}
For the Gaussian Orthogonal Ensemble (GOE), the corresponding expression is 
\begin{equation}
     \Delta_{3}(L) =  \frac{\ln L  - 0.0687 }{\pi^{2}}.
\end{equation}  
Integrable systems with Poisson statistics display weak rigidity, meaning $\Delta_{3}(L)$ grows linearly with $L$. In chaotic systems with Wigner–Dyson correlations, the spectral rigidity is enhanced and $\Delta_{3}(L)$ increases only logarithmically with $L$, reflecting stronger correlations over large ranges of the spectrum.\\

To determine the eigenvalues $E^{2}$ of eq (\ref{eq-eigen value eq}), we employ a pseudospectral method. For numerical implementation, the problem is restricted to a finite domain  $0 < y < y_{\max}, \,\,\, R_{\min} < R < R_{\max},$ and discretized on an $N \times N$ Chebyshev grid. We impose hard-wall Dirichlet boundary conditions at the edges of the domain, which requires the physical eigenfunctions to decay close to zero near the boundaries. This condition is satisfied by considering eigenvalues \(E^{2}\) that are much smaller than the value of the potential at the boundary.  For our purposes $R_{\max} = 10, \,\,\, R_{\min} = -10$ with N = 64. From the computed eigenvalues, we obtain the energy spectrum for each value of $k$.  We focus on eigenvalues with $4< k< 9$. Throughout the numerical analysis, without loss of generality, we have set the parameters to
\begin{equation}
    l=0.2,\,\,\,\lambda=1.5,\,\,\,
\end{equation}
 and the maximum of potential, $V_{max}$ varies depending on the specific choice of $\alpha$. It is to be noted that for each different value of $\alpha$ yields a different $y_{max}$. \\
To analyse the statistics, the spectrum is first unfolded such that the local mean spacing is normalized to unity. This yields the dimensionless nearest-neighbor spacings
\begin{equation}
    s_{i} = \frac{E_{i+1}-E_{i}}{\langle s \rangle},
\end{equation}
where \(\langle s \rangle\) denotes the local average level spacing. The resulting set of \(\{s_i\}\) is then used to construct a histogram, which represents the level spacing distribution. Thereafter, we further compute $\Delta_{3}(L)$ statistic characterizing spectral rigidity.\\

\begin{figure}[h]
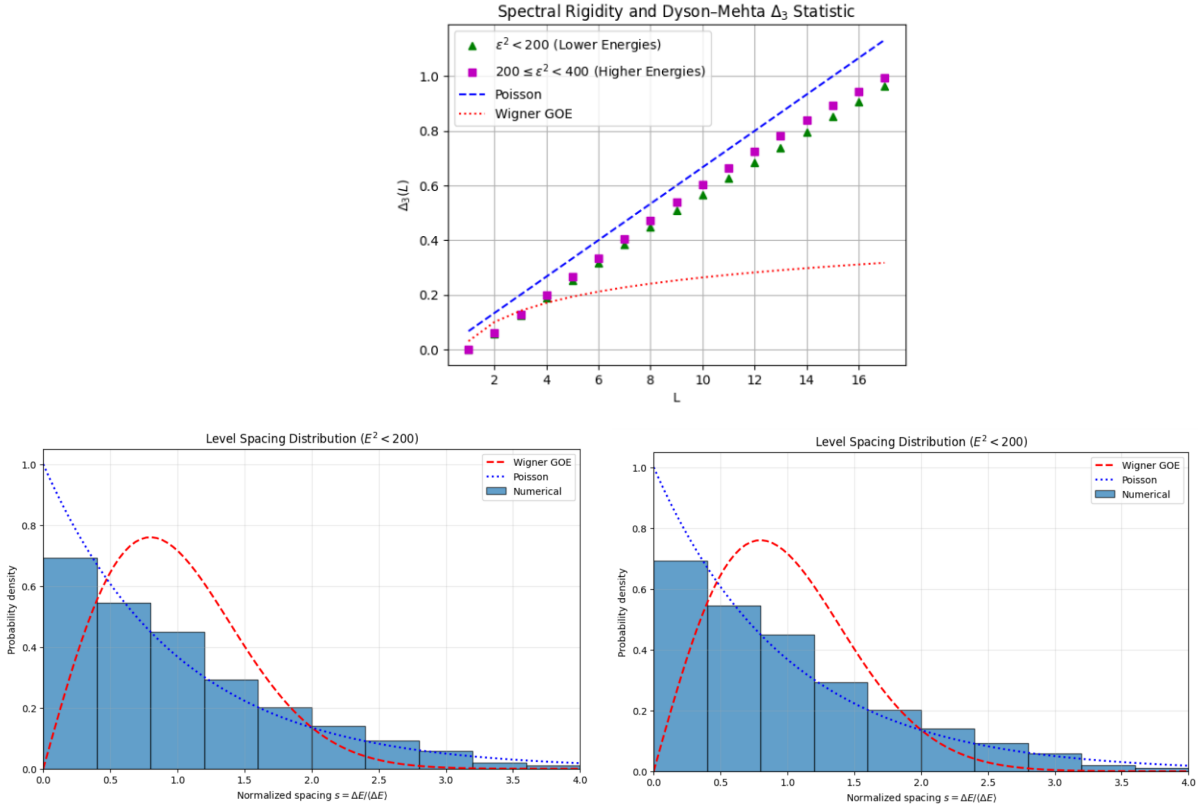

     \begin{subfigure}
        \centering
        \hspace{5cm}
    \includegraphics[width=0.4\linewidth]{new-new-DM-mycase-alpha-0.pdf}
    \end{subfigure}
    \newline
    \hspace{2cm}
    \begin{subfigure}
        \centering
    \includegraphics[width=0.45\linewidth]{new-new-LS-mycase-150-alpha-0.pdf}
    \end{subfigure}
    \begin{subfigure}
        \centering
    \includegraphics[width=0.45\linewidth]{new-new-LS-mycase-400-alpha-0.pdf}
    \end{subfigure}
    \caption{For the case $\alpha = 0$ with $V_{\max} = 400$
    having energy cutoffs $E^2 < 200$ and $E^2 < 400$, the upper panel presents the spectral rigidity and Dyson–Mehta statistics. The dashed blue and dotted red curves correspond to the Poisson and Wigner GOE predictions, respectively. The results show consistency with the Poisson distribution across both the low- and high-energy regimes.
In the lower panel, the level-spacing distribution in the AdS soliton is depicted. Here too, for both low and high energies, the spectrum exhibits level repulsion while maintaining an overall behavior that remains close to the Poisson distribution.}
    \label{alpha-0}
\end{figure}

Let us first consider the limit $\alpha \xrightarrow{}0$ which defines a background with no hair \cite{Anabalon:2016izw}. 
For $\alpha \xrightarrow{}0$ and $\lambda=\eta$ in the conformal factor, the metric becomes that of a conformal negatively curved spacetime without hairiness as reviewed in section \ref{brief-hairy-ads-soliton}. 
It is evident that there are no roots of $f(x)=0$ for $\alpha \approx 0,\,\,l=0.2,\,\,\lambda=1.5$ as $f(x) $ becomes a constant at $\alpha \xrightarrow{}0$. Thus, there is no upper bound on $x_{max}$ or $y_{max}$. For numerical purposes, without loss of generality, let's take $y_{max}=25$. Fig \ref{alpha-0} shows a Poisson-like distribution of both lower and higher energy eigenvalues in Dyson-Mehta statistics (upper panel) as well as in the histograms of level spacing (lower panel). This explains that the string probing the metric without hair acquires integrable dynamics.
\begin{table}[h]
\centering
\begin{tabular}{c|c|c|c|c|c|c}
\toprule
\textbf{Set} & \textbf{Parameters} & \multicolumn{2}{c|}{\textbf{DM range of}} & \multicolumn{2}{c|}{\textbf{LS range of}} & \textbf{Remarks} \\
             &                     & \textbf{GOE} & \textbf{Poisson}          & \textbf{GOE} & \textbf{Poisson}         & \\ 
\midrule
\multirow{5}{*}{1} 
  & $\alpha = -0.1$ 
   & $\,\,E^2< 150\,\,$  & $\,150 < E^2 < 400\,$ & $\,E^2< 150\,$ 
  & $\,150 < E^2 < 400\,$ 
  & See Fig \ref{fig 7} \\ 
  & $y_{\max} = 0.36$   &  &  &  &  &  \\
  & $V_{\max} = 400$  &  &  &  &  &  \\
\midrule
\multirow{5}{*}{2} 
  & $\alpha = -0.5$  
  & $E^2< 200$  & $200<E^2< 300$ & $E^2<200$
  & $ 200<E^2< 300$
  & See Fig \ref{set-2}\\
  & $y_{\max} = 0.30 $   &  &  &  &  &  \\
  & $V_{\max} = 600$  &  &  &  &  &  \\
\midrule\multirow{5}{*}{3} 
  & $\alpha = -1$   
  & $E^2< 210$  & $210<E^2< 400$ & $E^2< 210 $ 
  & $ 210<E^2< 400$
  & See Fig \ref{set-3} \\
  & $y_{\max} = 0.29 $   &  &  &  &  &  \\
  & $V_{\max} = 400$  &  &  &  &  &  \\
\bottomrule
\end{tabular}
\caption{Table showing parameters and their corresponding DM and LS ranges for chaos behaviour. Here, for all the sets, we choose: $\lambda=1.5,\,\,l=0.2$.}
\label{table-DM LS}
\end{table}
 \begin{figure}[h]
    \begin{subfigure}
        \centering
         \hspace{5cm}
    \includegraphics[width=0.45\linewidth]{new-DM.pdf}
    \end{subfigure}
    
    \begin{subfigure}
        \centering
    \includegraphics[width=0.45\linewidth]{new-new-chaos-150.pdf}
    \end{subfigure}
    \begin{subfigure}
        \centering
    \includegraphics[width=0.45\linewidth]{new-new-integrable-400.pdf}
    \end{subfigure}
    \caption{For the case $\alpha = -0.1$, the upper panel illustrates the spectral rigidity and the Dyson–Mehta $\Delta_{3}$ statistic. The spectrum agrees with the Wigner GOE behaviour at lower energies (green triangles, $E^2 < 150$), while at higher energies (magenta stars, $150 < E^2 < 400$), it transitions toward a Poisson-like distribution.
   In the lower panel, the level-spacing distribution is shown for the same hairy AdS soliton. At lower energies, the spectrum exhibits level repulsion characteristic of the Wigner GOE. However, this structure disappears at higher energies, where the distribution instead approaches that of Poisson statistics, indicating uncorrelated energy levels.}
    \label{fig 7}
\end{figure}

However, once the hair parameter $\alpha$ is switched on, the low-energy sector exhibits a distinct transition to Wigner-Dyson statistics, signaling the onset of quantum chaos characterized by level repulsion, while the high-energy part of the spectrum continues to follow Poisson statistics, thereby retaining integrability, see Fig \ref{fig 7}--Fig \ref{set-3}. A summary of the different parameter choices and the corresponding cases for each plot 
is provided in Table \ref{table-DM LS}.
From the plots, we can infer that chaos in the soliton background emerges predominantly in the IR region, whereas the UV sector remains integrable. Moreover, the extent of the chaotic regime is found to be sensitive to the magnitude of $|\alpha|$: for small values of $|\alpha|$, the chaotic window is narrow and confined to the lowest levels, but as $|\alpha|$ increases, the range of energy levels exhibiting chaotic behavior widens, indicating that the hair parameter effectively controls the strength and spread of chaos. At very large $|\alpha|$, the numerical analysis becomes unstable, suggesting either technical limitations or intrinsic constraints on the parameter space. We have deliberately chosen small values of $|\alpha|$, which would widen the energy window. Scenario of chaotic nature for lower energy eigenvalues and integrable nature for higher energy eigenvalues is similar to lower energy confinement and higher energy deconfinement with a dynamical critical phase transition point. A shift from chaos to integrability pertains along the IR to UV flow. Here, the critical phase transition point is governed by the combined effects of different parametric choices, in particular, the choice of $\alpha$. 
\\
\begin{figure}[h]
    \begin{subfigure}
      \centering
      \hspace{5cm}
      \includegraphics[width=0.45\linewidth]{new-new-corrected-DM-my-case-set-2.pdf}
    \end{subfigure}
    \newline
     \begin{subfigure}
      \centering
      \includegraphics[width=0.45\linewidth]{new-new-corrected-LS-150-case-set-2.pdf}
    \end{subfigure}
     \begin{subfigure}
      \centering
      \includegraphics[width=0.45\linewidth]{new-new-corrected-LS-400-case-set-2.pdf}
    \end{subfigure}
    \caption{For the case $\alpha = -0.5$, the upper panel illustrates the spectral rigidity and the Dyson–Mehta $\Delta_{3}$ statistic. The spectrum agrees with the Wigner GOE behaviour at lower energies (green triangles, $E^2 < 200$), while at higher energies (magenta stars, $200 < E^2 < 300$), it transitions toward a Poisson-like distribution.
   In the lower panel, the level-spacing distribution is shown for the same hairy AdS soliton. At lower energies, the spectrum exhibits level repulsion characteristic of the Wigner GOE. However, this structure disappears at higher energies, where the distribution instead approaches that of Poisson statistics, indicating uncorrelated energy levels.}
    \label{set-2}
\end{figure}
\begin{figure}[h]
    \begin{subfigure}
      \centering
      \hspace{5cm}
      \includegraphics[width=0.45\linewidth]{alpha_-1-DM-new.pdf}
    \end{subfigure}
    \newline
     \begin{subfigure}
      \centering
      \includegraphics[width=0.45\linewidth]{alpha_-1-LS-210.pdf}
    \end{subfigure}
     \begin{subfigure}
      \centering
      \includegraphics[width=0.45\linewidth]{alpha_-1-LS-400.pdf}
    \end{subfigure}
    \caption{
    For the case $\alpha = -1$, the upper panel illustrates the spectral rigidity and the Dyson–Mehta $\Delta_{3}$ statistic. The spectrum agrees with the Wigner GOE behaviour at lower energies (green triangles, $E^2 < 210$), while at higher energies (magenta stars, $210 < E^2 < 400$), it transitions toward a Poisson-like distribution.
   In the lower panel, the level-spacing distribution is shown for the same hairy AdS soliton. At lower energies, the spectrum exhibits level repulsion characteristic of the Wigner GOE. However, this structure disappears at higher energies, where the distribution instead approaches that of Poisson statistics, indicating uncorrelated energy levels.}
    \label{set-3}
\end{figure}
\section{Quantum Chaos via holography-based approach}\label{sec:quantumchaos}
From the statistical perspective, we obtained a clear sign of growing chaos in the dynamics of a probe string in our chosen solitonic background. This directly implies chaotic growth of boundary operators in the IR sector of the dual field theory. We will understand this phenomenon more rigorously in this section by investigating quantum chaos via two holography-based approaches, namely the entanglement wedge method and growth of out-of-time-ordered correlators in a shockwave analysis. In these methods, we will treat the solitonic tip analogously with the black hole horizon and check the nature of scrambling in our framework. We will also see an equivalence between these methods for a detailed analysis of the propagation of chaos in our solitonic system.
\subsection{Growth of out-of-time-ordered correlators }
\label{otoc}
We begin this section by constructing the double-sided extension of the hairy AdS soliton geometry and then perform the associated shockwave analysis. 
  This approach yields an emergent butterfly velocity and a quantity $\lambda_L = 2\pi / \beta$ which apparently resembles the Lyapunov exponent in a black hole system, but indeed characterizes an oscillatory growth of scrambling in the present solitonic system rather than a true exponential chaos. They also reflect the confinement/deconfinement transition.
\subsubsection{Double-sided geometry of hairy AdS soliton}
\label{double sided}
Here we will show the explicit construction of the double-sided Kruskal version of hairy AdS black hole given in (\ref{hairy black hole 5d}) by using a similar formalism presented in \cite{Jahnke:2017iwi} for generic anisotropic black holes. We assume near the horizon, the metric function $f(x)$ takes the form
\begin{align}
     G_{tt}(x) \equiv f(x) &= c_{0} (x- x_{h}),\\
    G_{xx}(x) &= \frac{c_{1}}{ x- x_{h}}
\end{align}
\begin{align}
    \beta = 4 \pi \sqrt{\frac{c_{1}}{c_{0}}}
\end{align}
We will work with the Kruskal coordinates to cover smoothly the two sides of the geometry. We introduce the following tortoise coordinate\begin{equation}
    x_*=-\int_0^x dx'\sqrt{\frac{G_{xx}(x')}{G_{tt}(x')}}=-\int_{0}^{x}\frac{\eta}{f(x')}dx'
\end{equation}
Now, we may see that we can write $\frac{\eta}{f(x')}=\sqrt{\frac{c_1}{c_0}}\frac{1}{x'-x_h}$. Let us use this in the expression of $x_*$ so that it gives us \begin{equation}
    x_*=\sqrt{\frac{c_1}{c_0}}\ln\left(\frac{x_h-x}{x_h}\right)=\frac{\beta}{4\pi}\ln\left(\frac{x_h-x}{x_h}\right)
\end{equation}
Then we define the Kruskal coordinates as follows 
\begin{align}
    UV= e^{\frac{4\pi}{\beta
    } x_{*}},\,\,\, U/V= e^{- \frac{4\pi}{\beta
    }t}\,.
\end{align} 

It is thus evident that the above expression will give $UV=0$ for $x=x_h$, i.e., at the horizon and $UV=1$ for $x=0$, i.e., at the black hole singularity. In this scenario, we can therefore write $x_*$ equivalently as 
\begin{equation}
    x_*=\frac{\beta}{4\pi}\left(\ln\Omega(x)+\ln f(x)-\ln{x_h}\right)
\end{equation}
Such formalism yields the double sided version of any anisotropic 5D black hole geometry as 
\begin{equation}
    ds^2=\frac{\beta^2}{4\pi^2}G_{tt}(U,V)\frac{1}{UV}dUdV+\sum_{i=1}^3G_{x_ix_i}(U,V)dx_i^2
\end{equation}Thus, for hairy black hole, we can write,
\begin{equation}
    ds^2=\Omega(U,V)\left[\frac{\beta^2}{4\pi^2UV}f(U,V)dUdV+\frac{1}{l^2}\sum_{i=1}^3dx_i^2\right]
\end{equation}
We define the tortoise coordinate as
\begin{align}
    x_{*}= \int_{0}^{x}\frac{\eta}{f(x')}dx' = \sqrt{\frac{c_{1}}{c_{0}}} \ln\left(\frac{x-x_{h}}{x_{h}}\right)
\end{align}
Then we define the Kruskal coordinates 
\begin{align}
    UV= e^{\frac{4 \pi}{\beta} x_{*}}, \,\,\,\, U/V= e^{\frac{4 \pi}{\beta}t}
\end{align}
which gives $UV=0$ for $x=x_{h}$ i.e., at horizon and $UV= -1$ for $x=0$ i.e. at the black hole singularity.\\ 
Then, the double-sided geometry of the 5D hairy black hole is given by
\begin{align}
    ds^2=\frac{\beta^2 \Omega(U,V) G_{tt}(U,V)}{4 \pi^2 UV}dUdV + \frac{\Omega(U,V)}{l^2}\sum_{i=1}^3dx_i^2
\end{align}
(until this, it defines a double-sided geometry with two asymptotically AdS regions connected by a wormhole.)
Next, we apply the double Wick rotation: $t \xrightarrow{} i\theta$ and $x_{1}\xrightarrow{}i\tau$. Then the Kruskal coordinates become 
\begin{align}
    \tilde{U}=e^{\frac{2\pi}{\beta}(x_*-i\theta)},~~\tilde{V}=e^{\frac{2\pi}{\beta}(x_*+i\theta)}
\end{align}
which gives $\tilde{U}\tilde{V}=UV$.
Then the metric takes the form after a double Wick rotation as
\begin{align}
    ds^2= \frac{\beta^2 \Omega(\tilde{U},\tilde{V}) G_{tt}(\tilde{U},\tilde{V})}{4 \pi^2 \tilde{U}\tilde{V}}d\tilde{U}d\tilde{V} + \frac{\Omega(\tilde{U},\tilde{V})}{l^2}\left(-d\tau^2+dx_2^2+dx_3^2\right)
    \label{double wick soliton}
\end{align}
One can easily see that the metric (\ref{double wick soliton}) can retrieve a hairy AdS soliton-like geometry if one use a suitable coordinate transformation. 
\subsubsection{Shockwave analysis}
\label{shockwave}
In this section, we introduce a small pulse of energy to the unperturbed metric (\ref{double wick soliton}) and analyse the consequences that follow. Before doing the calculation, we rename the coordinates $\hat{U},\hat{V}$ as simply $U,V$ for notational convenience in the metric \ref{double wick soliton}, so the unperturbed metric takes the form 
\begin{align}
     ds^2 = 2 A(U, V) dU dV + G_{ij}(U, V) dx^i dx^j
     \label{unperturbed-metric}
\end{align}
where, $A(U,V)= \frac{\beta^2 \Omega(U,V) G_{tt}(U,V)}{4 \pi^2 UV}$, and $G_{ij}= \frac{\Omega(U,V)}{l^2}$. Note that we have also renamed the coordinate $i \tau$ as $x_{1}$ only for purely notational purposes.  
The pulse is added from the boundary to the horizon on the left side of the geometry. We assume that the unperturbed metric (\ref{unperturbed-metric}) is the solution of the Einstein field equations with energy-momentum tensor 
\begin{align}
T^{matter}_{0}=2 T_{UV} dU dV + T_{UU}dU^2 + T_{VV}dV^2 + T_{ij}dx^i dx^j 
\end{align}
Next, we add a null pulse of energy located at $U=0$ and moving with the speed of light in V-direction. The pulse is located such that it remains unperturbed in the right exterior and only perturbation effects can be seen in the left exterior. Putting $V \xrightarrow{} V+ \Theta(U) \zeta(x^i)$, in the unperturbed metric, where $\Theta$ is the Heaviside function, we get
\begin{align}
    ds^2 
    &=  2 A(U, V + \Theta \zeta) dU (dV+\Theta \partial_i\zeta dx^i) + G_{ij}(U, V + \Theta \zeta) dx^i dx^j
\end{align}
and the energy-momentum tensor is given by
\begin{equation}\label{T_matter_wide}
\begin{split}
T^{\mathrm{matter}} &= 2\,T_{UV}\big(U,V+\Theta\zeta\big)\,dU\big(dV + \Theta\,\partial_{i}\zeta\,dx^{i}\big)
+ T_{UU}\big(U,V+\Theta\zeta\big)\,dU^{2} \\
&\quad + T_{VV}\big(U,V+\Theta\zeta\big)\,dV^{2}
+ T_{ij}\big(U,V+\Theta\zeta\big)\,dx^{i}\,dx^{j}\,.
\end{split}
\end{equation}
We define the following coordinates 
\begin{align}
    \bar{U} = U, ~~~\bar{V}= V+ \Theta(U), ~~~\bar{x}^i = x^i
\end{align}
In these coordinates, the metric becomes 
\begin{align}
ds^2= 2 \bar{A}d\bar{U}d\bar{V}- 2 \bar{A} \delta(\bar{U}) \zeta d\bar{U}^2 + \bar{G_{ij}}d\bar{x^i}d\bar{x^j}    
\end{align}
and the energy-momentum tensor becomes
\begin{align}
     T^{matter}= 2\left[\bar{T}_{\bar{U}\bar{V}} - \bar{T}_{\bar{V}\bar{V}}  \zeta  \delta(\bar{U}) \right] d\bar{U}  d\bar{V} + \bar{T}_{\bar{V}\bar{V}}  d\bar{V}^{2} + \bar{T}_{ij}  d\bar{x}^{i}  d\bar{x}^{j} \nonumber \\
+ \left[\bar{T}_{\bar{U}\bar{U}} + \bar{T}_{\bar{V}\bar{V}}  \zeta^{2}  \delta(\bar{U})^{2} - 2\bar{T}_{\bar{U}\bar{V}} \zeta  \delta(\bar{U}) \right] d\bar{U}^{2}
\end{align}
where the barred quantities are evaluated at the $(\bar{U},\bar{V}, \bar{x^{i}})$. We assume the localized perturbation and also assume the form of the backreacted energy-momentum tensor as \cite{Jahnke:2017iwi}
\begin{align}
    T^{shock}=  E e^{\frac{2i\pi \theta}{\beta} }\delta(\bar{U}) \delta(\bar{x}^i) d\bar{U}^2
\end{align}by using the analogy of black hole configuration and applying the analytic continuation $t\rightarrow i\theta$. With such assumption, we consider the temporal propagation of the shockwave in solitonic structure along the Euclidean time direction $\theta$.
The associated Einstein's equation reads as
\begin{align}
    R_{mn} - \frac{1}{2} G_{mn} R = 8\pi G_{\mathrm{N}} \left( T^{\mathrm{matter}}_{mn} + T^{\mathrm{shock}}_{mn} \right)
    \label{EEq}
\end{align}
where, $G_{N}$ is the $N$-dimensional Newton constant. Next, we do the following rescaling: 
\begin{equation}
    \zeta \xrightarrow{} \epsilon \zeta, ~~~ T^{shock} \xrightarrow{} \epsilon T^{shock} \label{scaling-eq}
\end{equation}
By doing this, we can recover the equations of motion for the unperturbed metric by setting $\epsilon= 0$ in \eqref{scaling-eq}. We assume that Einstein equations are satisfied for $\epsilon = 0$, and analyze the terms linear in $\epsilon$, we find that
\begin{equation}
\delta(U)\, G^{ij} \left( A\, \partial_i \partial_j - \frac{1}{2} G{_{ij}}_{,\ U V} \right) \zeta( x^i) = 8\pi G_N\, T^{\text{shock}}_{UU}
\label{shockequn}
\end{equation}
subject to the constraint \cite{Sfetsos:1994xa,Dray:1984ha}, 
\begin{equation}
A_{,V} = {G_{ij}}_{,V} = T^{\text{matter}}_{VV} = 0 \quad \text{at } U = 0.
\end{equation}
For double-sided hairy AdS soliton metric $G_{ij}(U,V)=\frac{\Omega(U,V)}{l^2}$. Then, 
\begin{align}
     \left[ \frac{ l^2A(U,V)}{\Omega(U,V)}(\partial_{x_2}^2+\partial_{x_3}^2)-\frac{1}{\Omega(U,V)}\frac{\partial^2\Omega(U,V)}{\partial V\partial U}\right]\zeta(\tau, x_2,x_3)= 4\pi G_{N}Ee^{\frac{2i\pi\theta}{\beta}}\delta(x_2)\delta(x_3)
\end{align}
The PDE becomes a 2D Helmholtz-like equation with constant coefficients:
\begin{equation}
    \Big[\frac{l^2 A_h}{\Omega_h}\nabla_\perp^2 - \frac{(\partial_U\partial_V\Omega)_h}{\Omega_h}\Big]\zeta(\tau,\mathbf{x}_\perp)
=4\pi G_N E e^{\frac{2i\pi\theta}{\beta}}\delta^{(2)}(\mathbf{x}_\perp),
\end{equation}
where $\nabla_\perp^2=\partial_{x_2}^2+\partial_{x_3}^2$,~~$\mathbf{x}_\perp=(x_2,x_3)$ \\
and, 
$$A_h \equiv A(U=0,V_0),\qquad \Omega_h\equiv\Omega(U=0,V_0),
\qquad (\partial_U\partial_V\Omega)_h\equiv\partial_U\partial_V\Omega|_{U=0,V_0}.$$
or, equivalently, 
\begin{align}
    \big(\nabla_\perp^2 - M^2\big)\zeta(\tau,\mathbf{x}_\perp)
= S(\tau)\,\delta^{(2)}(\mathbf{x}_\perp),
\end{align}
with 
\begin{align}
    M^2 \equiv \frac{(\partial_U\partial_V\Omega)_h}{l^2 A_h},
\qquad
S(\tau)\equiv \frac{4\pi G_N E e^{\frac{2i\pi\theta}{\beta}}\Omega_h}{l^2 A_h}.
\end{align}
Thus, the solution is given by
\begin{align}
    \zeta(\tau,r)
= -\,\frac{2 G_N E\,\Omega_h}{l^2 A_h}\,e^{\frac{2i\pi\theta}{\beta}}\,K_0(M r).
\end{align}
where,
$$r\equiv |\mathbf{x}_\perp|=\sqrt{x_2^2+x_3^2}$$ and  $K_0$ is the modified Bessel function of the second kind. This satisfies \begin{equation}(\nabla^2 - M^2)G(r)=-\delta^{2}(r)\,.\end{equation}
Thus, we can write $\zeta(\tau,r)\propto e^{\frac{2i\pi\theta}{\beta}}\,K_0(Mr)$. The large distance behavior of the modified Bessel function of second kind $K_{0}$ is $K_0(Mr)\sim \exp(-Mr)$. This can then be compared with
\begin{align}
    \zeta \sim \exp\!\Big[\lambda_{L}(i\theta-i\theta_* - r/v_B)\Big],\qquad \lambda_{L}=\frac{2\pi}{\beta}\,,
    \label{eq-lambda in shockwave}
\end{align}where $t$ being replaced with Euclidean time $\theta$. 
Therefore, we get the butterfly velocity of the spatial spread of scrambling as
\begin{align}
    v_B=\frac{\lambda_{L}}{M}=\frac{2\pi}{\beta}\frac{1}{M}
\end{align}
or , equivalently, 
\begin{align}
    v_B^2=\Big(\frac{2\pi}{\beta}\Big)^2\frac{1}{M^2}
=\frac{4\pi^2}{\beta^2}\,\frac{l^2 A_h}{(\partial_U\partial_V\Omega)_h}
\end{align}On the other hand, the growth of scrambling along the Euclidean time $\theta$ appears to be oscillatory. The quantity $\lambda_L=\frac{2\pi}{\beta}$ assumes exactly similar form as the Lyapunov exponent of exponential chaos in thermal systems, but in solitonic system, it characterizes the frequency of the oscillatory growth of scrambling along the Euclidean time. 
Now, the tip of the soliton geometry appears at $x_s=x_h$. Near the tip of our bulk geometry, 
\begin{align}
    x = x_s(1 + UV) \quad\Rightarrow\quad \partial_U x = x_s V,\ \partial_V x = x_s U,\ \partial_U\partial_V x = x_s
\end{align}
so, \begin{align}
    \partial_U\partial_V\Omega = \Omega'(x)\,\partial_U\partial_V x + \Omega''(x)\,\partial_U x\,\partial_V x
= x_s\Omega'(x) + x_s^2 UV\,\Omega''(x)
\end{align}
Assuming $U=0$ kills the second term, so
\begin{align}
    (\partial_U\partial_V\Omega)_h = x_s\,\Omega'(x)\Big|_{x=x_s}
\end{align}
Similarly, $G_{tt}(x)\simeq c_0(x-x_s)$ with $c_0=G_{tt}'(x_s)$, thus we obtain
\begin{align}
    A_s \;=\; \frac{\beta^2}{4\pi^2}\; \Omega_h\; c_0\; x_s
\end{align}
Hence, in the near-the-tip region, the butterfly velocity is given by
\begin{align}
    v_B = l\ \sqrt{\frac{\Omega(U,V)\,G'_{tt}(U,V)}{\partial_U\partial_V\Omega(U,V)}}\Big|_{UV=0}
    \label{vb-shock}
\end{align}which eventually yields
\begin{equation}
    v_B=\frac{l\sqrt{|\alpha|}}{60\sqrt{5}}\left[\frac{1}{x_s^{\frac{5}{2}}}\frac{(x_s^5-1)^2}{\sqrt{3x_s^5+2}}\right]
\end{equation}It is evident from the above analysis that the growth of scrambling with late time in such solitonic background is oscillatory but the spatial spread of the same attains exponential nature. Therefore, due to the solitonic structure of the bulk geometry, there occurs a decoupling between temporal and spatial growth of information scrambling in the boundary theory. This reveals that the system does not thermalise in time like a black hole but perceives efficient information transport covering exponentially more space. The rate of such spatial propagation of scrambling is given by the butterfly velocity (\ref{vb-shock}). 
In the expression (\ref{vb-shock}), the butterfly velocity attains zero value for $|\alpha|=0$. The metric (\ref{hairy AdS soliton 5d}) asymptotically reduces to a conformal flat spacetime in the limit $\alpha\rightarrow 0$ for any $\lambda$ or $\eta$. This shows that rate of the spatial propagation of scrambling vanishes at $\alpha=0$ which implies the integrability of the resulting conformal geometry 
with $\alpha=0$. It is to be noted that, for the soliton geometry, we treat the circumference $\beta$ of the Euclidean circle for a comparative analysis of thermodynamics of our chosen system \cite{Horowitz:1998ha}. In \cite{Maldacena:2015waa}, the chaos bound is studied only based on the thermal properties of the system, irrespective of the presence of black hole geometry.  We can therefore presume $\lambda_L=\frac{2\pi}{\beta_s}$ as a kinematic bound on the growth of scrambled dynamics of different observables in our background. Though there is no Hawking temperature, our geometry perceives certain thermal properties and hence the notion of scrambling in any thermal system may be used here. 
\subsection{Entanglement wedge reconstruction}
\label{EW method}
In parallel, chaotic dynamics can be probed through the construction of the entanglement wedge. The basic idea is to study how perturbations at the boundary affect the bulk extremal surfaces that define the entanglement entropy. In holography, the disruption of the entanglement wedge by an infalling perturbation provides a sharp diagnostic of information scrambling. As we obtained an exponential spatial deviation of gravitational perturbation during the shockwave analysis, it is straightforward to interpret it as a result of a nontrivial spatial entanglement and correlators. The aim in this section is to probe such spatial spread of scrambling by finding the size of the smallest boundary region whose entanglement wedge encloses the infalling particle. We will work in the same coordinate system as that used for the shockwave method. In the hairy AdS soliton, although no conventional horizon exists, the tip again plays the role of an effective obstruction for the entanglement wedge. This enables to define a “scrambling region” whose properties match those obtained from the shockwave analysis. From the dual field theory viewpoint, this construction offers an alternative perspective on the confinement/deconfinement dynamics and the onset of chaos, complementing the shockwave computation. 

First we consider the metric (\ref{hairy AdS soliton 5d}) of the 5D hairy AdS soliton and set the conformal boundary at \( x = 1 \). We consider a constant time slice and parametrize $x$ as $x=x(r)$, with $r=\sqrt{x_{2}^2+x_{3}^2}$ and $\theta$ remaining fixed. We can then write the induced metric on Ryu-Takayanagi(RT) surface parametrized by $(r, \phi)$ as
\begin{align}\label{induced metric}
    \gamma_{\alpha \beta}dx^{\alpha }dx^{\beta}=\Omega(x) \left[ \frac{ \lambda^2 x'^2}{f(x)} + \frac{1}{l^2} \right] dr^2 + \Omega(x) \frac{r^2}{l^2} d\phi^2
\end{align}
where $dx_{2}^2+dx_{3}^2=dr^2+r^2d\phi^2$.
Near the horizon, as the RT surface probes into the bulk (corresponding to late-time behaviour in the boundary theory), we approximate:
\begin{align}
    x(r) = x_s - \epsilon s(r)^2, \quad \epsilon \ll 1
\end{align}
where  $s(r)$ describes the surface profile.\\
Therefore, eq \ref{induced metric}
\begin{align}
     \gamma_{\alpha \beta}dx^{\alpha }dx^{\beta} = \Omega(x) \left[ \frac{4 \lambda^2 \epsilon^2 s^2 (s')^2}{f(x)} + \frac{1}{l^2} \right] dr^2 + \Omega(x) \frac{r^2}{l^2} d\phi^2
\end{align}
The entanglement entropy is proportional to the area of the RT surface, 
\begin{align}
    S_{EE}= 2\pi \int d^{d-1} y \sqrt{\gamma}
\end{align}
where $y$ are the coordinates of RT surface.
Next, approximating the induced metric up to the linear order in $\epsilon$: 
\begin{align}
     \gamma_{\alpha \beta}dx^{\alpha }dx^{\beta} \approx \Big[\frac{\Omega(x_{s})}{l^2}  - \Big(\frac{4 \Omega(x_{s})}{f'(x_{h})} \lambda^2 s'^2 + \frac{\Omega'(x_{s}) s^2}{l^2} \Big) \epsilon \Big]dr^2 + \Big(\frac{r^2\Omega(x_{s})}{l^2}  - \frac{\Omega'(x_{s})s^2 r^2}{l^2}\epsilon \Big)d\phi^2
\end{align}
The Lagrangian 
\begin{align}
    \mathcal{L} \sim \frac{\sqrt{\Omega(x_{s})}}{l}  - \Big(\frac{ 2l\sqrt{\Omega(x_{s})}}{f'(x_{s})} \lambda^2 s'^2 + \frac{\Omega'(x_{s})}{\sqrt{\Omega(x_{s})}} \frac{s^2}{l} \Big) \epsilon
\end{align}
Then the RT equation is given by varying w.r.t. $s(r)$ up to the linear order
\begin{align}
    s''(r) - \frac{1}{2l^2 \lambda^2} \frac{\Omega'(x_{s})}{\Omega(x_{s})}f'(x_{s})s(r)=0
\end{align}
On comparing with 
\begin{align}
    s(r) \sim \frac{e^{\tilde{\mu}r}}{r^\#}
\end{align}While looking at the metrics of hairy AdS black hole and hairy AdS soliton, it is evident that the function $f(x)$ is the same for both of them. So, in case of hairy black hole, the $f(x)=0$ at the horizon whereas $f(x)=0$ at $x=x_s$, say, for hairy AdS soliton where the geometry now gets smoothened due to the double Wick rotation. Thus, one can consider $x_s=x_h$.
Therefore, the butterfly velocity is 
\begin{align}
    v_{B} &\equiv \frac{2\pi}{\beta\tilde{\mu}}
          = \frac{2 \sqrt{2} \pi l \lambda \sqrt{\Omega(x_{s})}}{\beta \sqrt{\Omega'(x_{s})f'(x_{s})}}
          \label{vB-EWedge}
\end{align}At this stage, it is straightforward to compare the butterfly velocity obtained in this section with that computed through the exponential growth of the OTOCs in section \ref{otoc}. We can see that the equivalence between these two holography-based methods for hairy AdS soliton emerges when the compactification radius $\beta$ along $\theta$ direction takes the form
\begin{equation}
    \beta_s=\left( \frac{|\alpha|}{144\sqrt{2}\pi\lambda^4\Omega^{\frac{3}{2}}(x_s)}\right)^{-1}\label{temp of hairy soliton}
\end{equation}
This plays significant role in shaping thermodynamics of the theory associated to the chosen hairy AdS soliton background. 
In section \ref{sec:critical phase transition}, we will further undergo an approximate evaluation of the explicit dependency of the butterfly velocity on the hairy parameter $\alpha$ by substituting the form of $x_s$ as a function of $\alpha$. This will help us understanding the critical phase transition scenario of hairy AdS soliton backgrounds. 
 \\

\section{ Comments on Phase transitions} \label{sec:critical phase transition}
In \cite{Anabalon:2016izw}, while studying the phase transition via the free energy $\Delta F$, they found the critical temp $T_{c}= \frac{1}{L_{s}}$, $L_s$ being the radius of compactification along the direction $x_1\rightarrow i\tau$. It is crucial to note that even though the $x_{h}=x_{s}$, but their physical meaning is subject to the conformal factor $\Omega$. In this section we will see that the hairy parameter $\alpha$ plays a crucial role in the critical phase transition in the background under consideration by studying the variation of the butterfly velocity as a function of $\alpha$. \\


{The tip $x=x_s$ of our desired background appears exactly where the horizon of the hairy AdS black hole lies, i.e., at $x=x_s=x_h$ where $f(x)|_{x=x_s}=0$. This gives
\begin{equation}
    \alpha(x_s)=-\frac{90000}{l^2}\left(x_s^{10}-6x_s^5+30\log{x_s}+3+\frac{2}{x_s^5}\right)^{-1}\label{alpha in xs}\end{equation}From this expression, it is evident that the butterfly velocity obtained through entanglement wedge method and shockwave analysis depend nontrivially on the intrinsic hairy parameter $\alpha$ of the hairy AdS soliton geometry. Using this expression, we can write the inverse function $x_s=x_s(\alpha)$ and substitute it in the expression of the butterfly velocity. However, equation (\ref{alpha in xs}) is not solvable in closed form, as it is transcendental due to the logarithmic term mixed with polynomial and rational functions of $x_s$. Hence, we will undergo a numerical analysis to obtain the behaviour of the butterfly velocity $v_B$ and  $\lambda_L=\frac{2\pi}{\beta}$ with the variation of the parameter $\alpha$.
\begin{figure}
    \centering
    \includegraphics[width=0.45\linewidth]{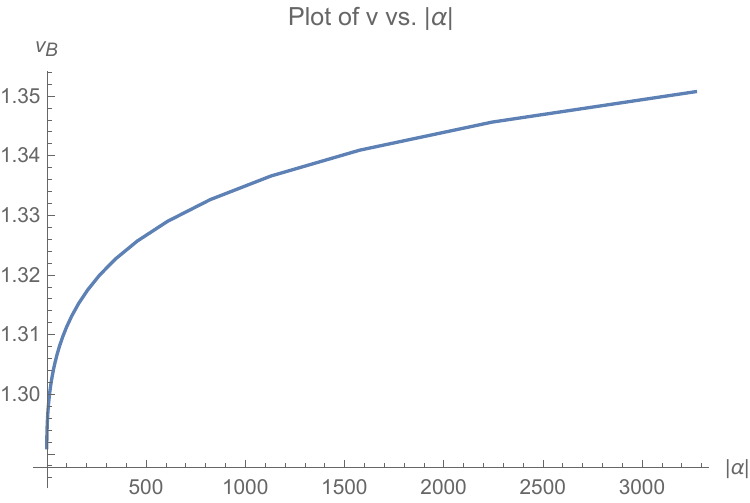}
    \caption {Numerical plot of $v_B$ with the variation of the hairy parameter $|\alpha|$ by setting domain of $x_s$ from $1.5$ to $5$.}
    \label{new-vB-plot}
\end{figure}Figure \ref{new-vB-plot} shows that the butterfly velocity is zero for $|\alpha|=0$ and attains finite non-zero values as soon as the hairy effect starts to grow. However, $v_B$ grows very slowly with $|\alpha|$. 
This implies that the scrambling of information inside the hairy AdS soliton bulk is almost localized and it is caused due to the presence of finite scalar hair in the background. Observations in this section are consistent with those obtained via the statistical approaches in section \ref{numerical-apprach} where a chaotic window appears in the lower energy eigenvalue spectrum only when the hair parameter $|\alpha|$ assumes finite values and it extends more with the increase in $|\alpha|$. From the perspective of the dual field theory with chemical potential, our result seems to be consistent with the phase diagram of chaos obtained in \cite{Akutagawa:2018yoe} for a linear sigma model as well as holographic D4/D6 brane model of a QCD-like theory with broken U(1) gauge symmetry. Moreover, our observation also resembles the slow and logarithmic scrambling occurring in quantum many-body localized (MBL) systems \cite{Swingle:2016jdj,B_lter_2022,chen2016universallogarithmicscramblingbody}.
Our observation is also suggestive toward a possible interplay between the dynamical transition from integrability to chaos and the phase transition between insulator to superconductor in a solitonic geometry as both of these are triggered by the appearance of a finite scalar hair. 

We can also probe the black hole to soliton phase transition from the expression of $\beta_s$. The Hawking temperature  of the hairy AdS black hole analytically constructed in \cite{Anabalon:2016izw} is given by
\begin{equation}
    T=\frac{1}{\beta_b}=\frac{|\alpha|}{288\pi\eta^4|\Omega(x_h)|^{\frac{3}{2}}}
\end{equation}Comparing this with (\ref{temp of hairy soliton}), we get a ratio 
\begin{equation}
    \frac{\beta_s}{\beta_b}\sim\left(\frac{\eta}{\lambda}\right)^4\,,
\end{equation}which implies that soliton to black hole phase transition, or, in other words, confinement/decon-\\finement phase transition occurs when $\eta\sim\lambda$ and the critical temperature depends nontrivially on the hairy parameter $\alpha$. This is consistent with the analysis of such phase transition from the notion of the thermodynamics of the hairy black hole in \cite{Banerjee:2007by, Anabalon:2016izw,Anabalon:2022ksf}. It is worth mentioning that  the critical temperature $T_c$ at the phase transition point crucially depends on $|\alpha|$. The metric with $f(x,\alpha)|_{|\alpha|=0}$ happens to be a zero temperature solution of the associated bulk gravity action. The thermal properties are acquired only for nonzero $|\alpha|$ values. The phase diagram depicted in the left plot of Fig \ref{new-Tc-plot} exhibits a very slow but monotonic increase in the critical temperature $T_c$ with the growth of hairiness in the bulk. Therefore, the confinement/deconfinement phase transition becomes dynamical due to the presence of finite bulk scalar hair. The region below the red line captures the confined sector and, above the red line, it defines deconfined sector in the phase diagram. Subsequently, the $|\alpha|$ dependency of the compactification radius $\beta_s$ reduces the $\lambda_L$ to 
\begin{equation}
    \lambda_{L}=\frac{2\pi}{\beta_{s}}=\frac{g(\alpha)}{72\sqrt{2}\lambda},~~g(\alpha)=\frac{|\alpha|}{\Omega^{\frac{3}{2}}(x_s(\alpha))}
\end{equation}It is obvious from the expression of $\beta_s$ that $\lambda_L$ is essentially zero when $|\alpha|=0$. Henceforth, it supports the concurrence of dynamical transition from integrability to chaos and the well-known insulator/superconductor phase transition. Nevertheless, the growth of $\lambda_L$ with $|\alpha|$ is very slow and subjective to quite higher values of $|\alpha|$ as given in the right plot in Figure. \ref{new-Tc-plot}.
\begin{figure}
    \centering
    \includegraphics[width=0.45\linewidth]{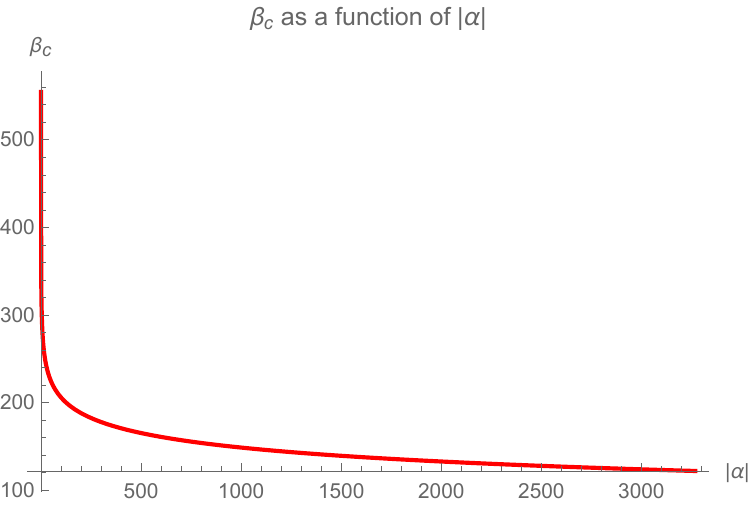}
    \includegraphics[width=0.45\linewidth]{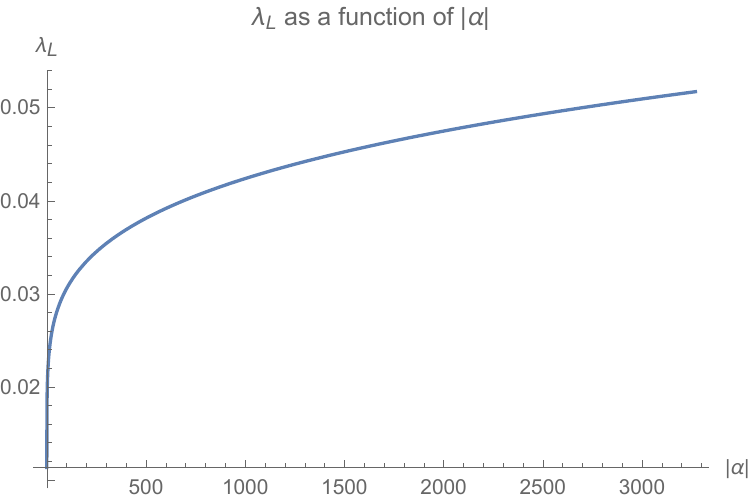}
    \caption{ The figure in the left shows the dynamical nature of the confinement/deconfinement phase transition on the basis of $\beta_s=\beta_b=\beta_c$, say, of the soliton geometry at the critical phase transition point. We choose $\eta\sim\lambda\sim1$ at the critical point. $\beta_c$ monotonically decreases with $\alpha$. The region below the red line is deconfined(black hole) and that above the red line is confined(soliton). The right figure shows the corresponding variation of $\lambda_L$.}
    \label{new-Tc-plot}
\end{figure} 
\section{Conclusions and Discussions} \label{conclusion}
We investigate the signatures of classical and quantum chaos in a suitable analytical five-dimensional class of hairy AdS soliton background. 
We probe classical chaos in our chosen background by studying the dynamics of a free-falling particle and by means of the Poincar\'e section and Lyapunov exponent of a fundamental closed string . On the other hand, we analyse the quantum nature of chaos and information scrambling through a couple of seemingly  different approaches, to name, spectral analysis and holography-based approaches. During spectral analysis, we capture quantum chaos using level-space repulsion and Dyson-Mehta statistical distribution of the quantized string spectrum. The holography-based approaches include an investigation of the scrambled growth of out-of-time-ordered correlators and the information scrambling in an entanglement wedge constructed inside the bulk.

 Classical chaos is found to be suppressed while probing it via the Rindler momentum of the particle motion. This happens due to the absence of a proper horizon in the soliton geometry. We further explore the chaotic motion of a fundamental probe classical string in the hairy AdS soliton background under consideration. It generates a gradual disruption of the Poincar\'e section for higher energies of the string states, leading to the positive (maximum) Lyapunov exponent and hence indicates for the chaotic dynamics. 
 We then undergo a mini superspace quantization of the string spectrum and explore the statistical spectral approaches. Both the studies of level-spacing repulsion and Dyson-Mehta statistics result in a Wigner-GOE distribution of the lower energy spectrum while the higher energy spectrum follows Poisson distribution. However, when $|\alpha|=0$, the distribution is Poisson-like irrespective of the energy eigenvalues. It immediately reveals that, at quantum level, the background without the scalar hair supports integrable string dynamics. As soon as the effect of the scalar hair is turned on, $|\alpha|$ assumes finite value and triggers the appearance of chaotic distribution in the lower energy eigenvalue sector. Such chaotic window in the string spectrum gradually spreads through higher energies with an increase in the values of $|\alpha|$.  Similarly as suggested in \cite{Shukla:2024wsu}, our study of chaotic behaviour of closed string in a hairy AdS soliton background thus implies a chaotic growth of the glueball operator spectrum  in the IR sector of confining theories with finite chemical potential.
To gain deeper holographic insights, we finally employ the holography-based approaches of information scrambling. Here, we treat the horizon of the hairy AdS black hole and the tip of the hairy AdS soliton geometry on equal footing as both of these ensure the IR regime of the corresponding dual theories. For the derivation of OTOC, we carry out a shockwave analysis followed by constructing a double sided Kruskal version of our soliton geometry in analogy with that of a black hole. Introducing a gravitational shockwave along one of the Kruskal directions, we solve the associated perturbed Einstein's equation of motion. We achieve an oscillatory growth of scrambling along the Euclidean time direction whereas the spatial scrambling still grows exponentially. We deduce the near-the-tip butterfly velocity of the spatial growth of scrambling as well as the frequency of the temporal oscillatory scrambling. A positive and real near-tip butterfly velocity for finite $|\alpha|$ suggests a scrambling of the glueball operator spectrum in the IR phases of dual confining boundary theory with finite chemical potential.
We also find an emerging equivalence between the butterfly velocities obtained from OTOC analysis and entanglement wedge method for a typical expression of the compactification radius $\beta_s$ along the Euclidean time direction. The expression of $\beta_s$ in our analysis governs the critical soliton/black hole phase transition (confinement/deconfinement phase transition in dual boundary field theory) consistently with that studied using black hole thermodynamics. It is worth noting that the finite scalar hair yields a dynamical nature of the critical phase transition. Added to this, we analyze the behaviour of the butterfly velocity and the chaos bound with the hair parameter $|\alpha|$. We find an interesting interplay between the transition from integrable to chaotic dynamics and the insulator/superconductor phase transition, as both of these are initiated by the presence of the finite hairy effect. The holographic phases with no hair are insulators and those with finite hair become superconductor. As scrambling grows with  $|\alpha|$, we speculate an underlying connection between these two kinds of transitions. However, the growth of scrambling with $|\alpha|$ is found be very slow and almost localized which may be thought as somewhat similar to the slow logarithmic growth of scrambling in some typical quantum many-body localized systems.\\ 

Although, preliminarily, the absence of a proper horizon suppresses the chaotic motion of a classical particle, we classically diagnose a chaotic evolution of the Poincar\'e section and Lyapunov exponent for higher energy states of a probe string, while mimicking the tip of our geometry as the horizon of the corresponding black hole. This analogy is heuristic, as the soliton geometry lacks a genuine event horizon, but it facilitates the application of methods developed for black holes. 
Furthermore, the quantum analysis of the string spectrum somewhat contradicts the results achieved from the classical string perspective. It explains that quantum effects induce some inherent or ``hidden" properties in the chaotic string dynamics in the lower eigenvalue spectrum, which we can comprehend using quantum eigenvalue distributions, but not via classical analysis.

Such issues thus require more precise and advanced investigation. An immediate and potential follow-up from our study is to construct the thermofield double states and explicitly derive the scrambled OTOCs for a hard-wall QCD-like gapped theory with finite chemical potential that may be considered as a competent candidate dual theory for the hairy AdS soliton background. This will help us understand the extent of consistently applying the black hole analogy in studying scrambling of information in such system. It would also be fascinating to conduct a similar analysis as above for quantum corrected AdS solitons \cite{Thompson:2024vuj} and observe the effect of quantum correction in the scrambling of information dynamics and associated phase transition scenario in AdS soliton backgrounds with or without scalar hair. One may also check the growth of quantum complexity in different glueball operators in hard wall holographic QCD theories (with or without chemical potential) in the IR regime. This will help sharpen our insights of different phase transitions in context of solitonic geometries on the basis of the growth of complexity. Furthermore, it will be crucial to search for stronger and more concrete understanding on the concurrence of integrability to chaos dynamical transition and phase transition between insulator and superconductor phases of holographic matter. We will come back with some of these understandings in near future.

\section*{Acknowledgments}
BS acknowledges support from the IIT Kharagpur Institute Fellowship and thanks Prof. Shiraz Minwalla for fruitful suggestions on the calculations of the shockwave analysis in the AdS solitonic geometries during his visit to IIT Kharagpur. BS also thanks Bhaskar Shukla for some discussion during the early stages of this manuscript. AC would like to thank the  Research project supported by the program ``Excellence initiative – research university" for the AGH University under the project IDUB 501.696.7996 for providing funds to carry out the above work. AC would like to thank Prof. Niko Jokela for extensive and insightful discussions and suggestions on different kinds of phase transitions in confining backgrounds during a visit to the Helsinki Institute of Physics(HIP), University of Helsinki. AC would like to acknowledge the Miniatura-8 short term research visit grant with grant number 18.18.220.09030 by National Science Centre(NCN), Ministry of Education and Science, Government of Poland for supporting the visit to HIP, University of Helsinki. AC would like to thank Prof. Julian Sonner for his valuable suggestions on the information scrambling and thermalization in solitonic geometry during the workshop ``Advanced school on complexity, chaos and integrability", University of Warsaw.


 \vspace{10pt}

\bibliographystyle{JHEP}
\bibliography{Soliton.bib}

\end{document}